 \documentclass[manuscript, screen]{acmart}

\AtBeginDocument{%
  \providecommand\BibTeX{{%
    \normalfont B\kern-0.5em{\scshape i\kern-0.25em b}\kern-0.8em\TeX}}}

\copyrightyear{2022} 
\acmYear{2022} 
\setcopyright{acmcopyright}\acmConference[CHI '22]{CHI Conference on Human Factors in Computing Systems}{May 8--13, 2021}{New Orleans, USA}
\acmBooktitle{CHI Conference on Human Factors in Computing Systems (CHI '22), April 30 -- May 6 2022, New Orleans, LA, USA}
\acmPrice{15.00}
\acmISBN{}

\usepackage{subfigure}
\usepackage{multirow}
\usepackage{xcolor}
\usepackage{microtype}
\usepackage{array}
\usepackage{colortbl}
\usepackage{tabularx}
\newcolumntype{\$}{>{\global\let\currentrowstyle\relax}}
\newcolumntype{^}{>{\currentrowstyle}}

\usepackage{hyperref}

\def\markup{0} 
\if\markup 1
\usepackage{soul}
\newcommand{\rv}[1]{{\leavevmode\color{blue}#1}}
\else
\newcommand{\rv}[1]{#1}
\newcommand{\st}[1]{}
\fi

\begin{document}

\title{``I Don't Want People to Look At Me Differently''} 
\subtitle{Designing User-Defined Above-the-Neck Gestures for People with Upper Body Motor Impairments}


\author{Xuan Zhao}
\affiliation{%
  \institution{School of Information, Rochester Institute of Technology}
  \city{Rochester}
  \country{USA}
}
\email{xz7320@rit.edu}

\author{Mingming Fan}
\authornote{Corresponding author}
\affiliation{%
  \institution{Computational Media and Arts Thrust, The Hong Kong University of Science and Technology (Guangzhou)}
  \city{Guangzhou}
  \country{China}
}
\affiliation{%
  \institution{Division of Integrative Systems and Design \& Department of Computer Science and Engineering, The Hong Kong University of Science and Technology}
  \country{Hong Kong SAR, China}
}
\email{mingmingfan@ust.hk}

\author{Teng Han}
\affiliation{%
  \institution{Institute of Software, Chinese Academy of Sciences}
  \city{Beijing}
  \country{China}
}
\email{hanteng@iscas.ac.cn}



\renewcommand{\shortauthors}{Zhao, Fan, and Han}

\begin{abstract}

Recent research proposed eyelid gestures for people with upper-body motor impairments (UMI) to interact with smartphones without finger touch. However, such eyelid gestures were designed by researchers. It remains unknown what eyelid gestures people with UMI would want and be able to perform. Moreover, other above-the-neck body parts (e.g., mouth, head) could be used to form more gestures. We conducted a user study in which 17 people with UMI designed above-the-neck gestures for 26 common commands on smartphones. We collected a total of 442 user-defined gestures involving the eyes, the mouth, and the head. Participants were more likely to make gestures with their eyes and preferred gestures that were simple, easy-to-remember, and less likely to draw attention from others. We further conducted a survey (N=24) to validate the usability and acceptance of these user-defined gestures. Results show that user-defined gestures were acceptable to both people with and without motor impairments.

\end{abstract}



\maketitle

\section{Introduction}
People with upper body motor impairments have difficulty touching an on-screen target accurately with fingers due to tremors, muscular dystrophy, or loss of arms~\cite{10.1145/3173574.3174094,10.1145/3236112.3236116}. 
They were found to have difficulty entering and correcting texts, grabbing and lifting the phone, making multi-touch input, pressing physical buttons, and so on, especially outside of home~\cite{10.1145/2661334.2661372}.
While voice-based interfaces, such as Siri~\cite{10.1145/3151470.3151476}, could be an alternative method, they suffer from low input speed and accuracy~\cite{10.1145/3151470.3151476,10.1145/2661334.2661372} and may raise privacy and social acceptance concerns when used in public~\cite{10.1145/2702123.2702188,10.1145/2661334.2661372}. 

Recently researchers proposed eyelid gestures for people with motor impairments to subtly interact with mobile devices without finger touch or drawing others' attention~\cite{10.1145/3373625.3416987,fan2021eyelidCAM}. These eyelid gestures, though followed design principles, were created without involving people with motor impairments in the design process. Consequently, it remains questionable whether these gestures are the ones people with motor impairments preferred in the first place. Indeed, the participants with motor impairments in their study~\cite{10.1145/3373625.3416987} also suggested other eyelid gestures and expressed a desire to design \textit{their own} eyelid gestures. Thus, there is a need to involve people with motor impairments to design gestures that will be used by them. In the meantime, prior research demonstrated that allowing users to define their preferred gestures would uncover more representative and preferred gestures~\cite{10.1145/1518701.1518866,10.1145/2468356.2468527,10.1145/1978942.1978971}.

Motivated by this need and prior success of designing user-defined gestures in other contexts, we sought to engage people with motor impairments to design eyelid gestures they prefer. What's more, as recent research demonstrated the promise of gaze and head pose for hands-free interaction in addition to eyelid gestures~\cite{nukarinen2016evaluation,sidenmark2019eye,kyto2018pinpointing,yan2018headgesture}, we extended the design space of user-defined gestures by inviting people with motor impairments to design \textit{above-the-neck gestures} that include eyelids, gaze, mouth, and head. 

We conducted an online user study, in which 17 participants with various upper body motor impairments designed above-the-neck gestures to complete 26 tasks that were commonly performed on mobile devices. These tasks included general commands (e.g., tap and swipe), app-related commands (e.g., open an app or a container with the app), and physical button-related commands (e.g., volume up and down). During the study, participants first watched a video clip explaining each task and its effect on a smartphone and then had time to design and perform an above-the-neck gesture. Afterward\st{s}, participants rated the goodness, ease, and social acceptance of the gestures they just created. Finally, they were interviewed to provide feedback on the gestures. 
We collected a total of 442 user-defined gestures. 
Our results show that participants preferred to use gestures that were simple, easy-to-remember, and less attention-demanding. Based on all the gestures obtained and the rating and frequency of use of each gesture, we assigned each command the most appropriate gesture or gesture. 

To validate the usability and acceptance of these user-defined gestures, we conducted an online survey that asked participants to select the most appropriate gesture from three candidates to complete each of the 
that we did previously in the user study on mobile devices. These candidate options were chosen from the most frequently mentioned gestures from the user study. Results show that our gesture set was well accepted and recognized by people with and without motor impairments.  In sum, we make the following contributions in this work: 
\begin{itemize}
\item We present a set of user-defined above-the-neck gestures based on the gestures designed by people with upper body motor impairments to complete common interactions on mobile devices;
\item We show that these user-defined above-the-neck gestures are largely preferred by people with and without motor impairment.
\end{itemize}


\section{Background and Related Works}
Our work is informed by prior work on \textit{interaction techniques for people with motor impairments} and \textit{user-defined gesture designs}.

\subsection{Interaction Techniques for People with Motor Impairments}
Brain\rv{-}computer interfaces (BCIs) sense brain signals for people with motor impairments to communicate with the environment or control computer systems without using hands (e.g.,~\cite{10.3389/fnhum.2018.00014,corralejo2014p300}). 
However, BCIs often need long periods of training for users to control their brain rhythms well~\cite{pires2012evaluation}, and people with motor impairments were reported to be concerned about fatigue, concentration\rv{,} and also social acceptance~\cite{taherian2017we,blain2012barriers}.

Gesture-based interactions have been investigated as an alternative approach. Ascari et al.~\cite{10.1145/3408300} proposed two machine learning approaches to recognize \textit{hand gestures} for people with motor impairments to interact with computers. However, these approaches are not feasible for people with upper body impairments who could not use their hands freely. 
To overcome the limitations of body and hand gestures for people with upper body impairments, researchers investigated \textit{eye-based} interactions.  

Among all eye-based gestures, \textit{blink} was probably the most widely studied for people with motor impairments. Earlier work used EOG sensors to detect blink to trigger computer commands~\cite{kaufman1993}. 
The \textit{duration} of blink was also utilized as additional input information. For example, \rv{a} long blink was detected and used to stop a moving onscreen target~\cite{Heikkila2012}.
What's more, blink was also used along with \textit{eye movements} to trigger mouse click~\cite{Kwon1999}. 
Blink was also used along with \rv{the} head motion. For example, the frequency of blink combined with \textit{head motion} \st{were}\rv{was} used to infer five activities, including reading, talking, watching TV, math problem solving, and sawing~\cite{Ishimaru2014}.


In addition to blink, \textit{wink} was used for people with motor impairments. Shaw et al. constructed a prototype to detect the open and close states of each eye and used such information to infer three simple eyelid gestures: blink, wink the left eye, and wink the right eye~\cite{Shaw1990}.
Similarly, Zhang et al. proposed an approach to combine blinks and winks with \st{with }gaze direction to type characters~\cite{Zhang:2017:SGG:3025453.3025790}. Recently, Fan et al. took a step further to investigate the design space of \textit{eyelid gestures} and proposed an algorithm to detect nine eyelid gestures on smartphones for people with motor impairments~\cite{10.1145/3422852.3423479,10.1145/3373625.3416987}.
These eyelid gestures were designed based on eyelid states, in which two eyelids could be in, and the possible parameters that humans can control, such as the duration of closing or opening an eyelid and the sequence. 
Although the design of eyelid gestures followed a set of design principles, these gestures were designed by the researchers who did not have motor impairments themselves, and the design process did not involve people with motor impairments in the loop. Consequently, it remains unknown \textit{whether these eyelid gestures were the ones that people with motor impairments preferred in the first place}. In fact, participants with motor impairments in their study could not well perform some eyelid gestures well, proposed new eyelid gestures\rv{,} and expressed the desire to design their own gestures. Motivated by this need, we seek to explore user-defined eyelid gestures that people with motor impairments would want to create and use. 

Other body parts, such as the head, have also been used to extend the interaction for people with motor impairments. Kyto et al.~\cite{kyto2018pinpointing} compared eyes and head\rv{-}based interaction techniques for wearable AR and found that the head-based interactions caused less error than eye-based ones. What's more, they found the combination of eye and head resulted in a faster selection time. Sidenmark and Gellersen~\cite{sidenmark2019eye} studied the coordination of eye gaze and head movement and found this approach was preferred by the majority of the participants because they felt better control and less distracted.
Similarly, gaze and head turn were also combined to facilitate the control of onscreen targets~\cite{nukarinen2016evaluation}.
Inspired by this line of work that shows the advantage of combining head motion with eye motions, we extend our exploration to include above-the-neck body parts, including both eyes, head and mouth, to allow people with motor impairments to better design a richer set of user-defined gestures.  

\subsection{User-Defined Gesture Designs}


User-defined gestures have been investigated by researchers in various contexts~\cite{10.1145/1518701.1518866,10.1145/2468356.2468527,10.1145/2166966.2166984,10.1145/2702613.2732747,10.1145/3334480.3382883,10.1145/3385959.3422694,10.1145/1978942.1978971,10.1145/3365610.3365625,10.1145/2598153.2598184}. Wobbrock et al.~\cite{10.1145/1518701.1518866} studied user-defined gestures for multi-touch surface computing, such as tabletop. They investigated what kind of hand gestures non-technical users would like to create and use by asking the participants to create gestures for 27 referents with one hand and two hands. Wobbrock et al.~\cite{10.1145/1518701.1518866} also designed gestures for the 27 referents on their own and compared the gestures created by them with the ones created by the users. They found that they created far fewer gestures than participants, and many of the gestures they created were never tried by users. Kurdyukova et al.~\cite{10.1145/2166966.2166984} studied the user-defined iPad gestures to transfer data between two displays, including multi-touch gestures, spatial gestures, and direct contact gestures. Piumsomboon et al.~\cite{10.1145/2468356.2468527} worked on user-defined gestures for AR and asked people to perform hand gestures with a tabletop AR setting while Lee et al.~\cite{10.1145/2702613.2732747}utilized an augmented virtual mirror interface as a public information display. Dong et al.~\cite{10.1145/3334480.3382883,10.1145/3385959.3422694}worked on the user-defined surface and motion gestures for mobile AR applications. In the work of Ruiz et al.~\cite{10.1145/1978942.1978971}, they utilized the user-defined method to develop the motion gesture set for mobile interaction. Weidner and Broll~\cite{10.1145/3365610.3365625} proposed user-defined hand gestures for interacting with in-car user interfaces, and Troiano et al.~\cite{10.1145/2598153.2598184} presented for interacting with elastic, deformable displays. 
These user-defined methods motivated our research. Specifically, our research investigates what upper-neck gestures people with motor impairments would like to create and how they would want to use such gestures to accomplish tasks on their touch-screen mobile devices. 


\section{Method}
The goal of this IRB-approved study was to gather user-defined above-the-neck gestures for common tasks on mobile devices from people with motor impairments and then identify the common user-defined gestures for each task. 

\begin{table*}[htb!]
    \caption{Participants' demographic information}
      \label{participant}
    \Description{Table demonstrates the age, sex, location, and prior device usage experiences, such as computer, android phone, iPad, and iPhone of 16 participants each.}
    \begin{tabular}{c|c|c|c}
    \hline
     \rowcolor[gray]{0.9}ID & Age & Gender & Motor impairments \\
    \hline 
    P1 &  32 & Male & Spinal cord injuries, wheelchair user\\ \hline
    P2 &  25 & Male & Cerebral palsy, shaking hands and hard to control hand movements\\ \hline
    P3 &  30 & Male & Loss or injury of limbs loss of both of arms\\ \hline
    P4 & 19 & Female & Cerebral palsy, shaking hands and hard to control hand movements\\ \hline
    P5 &   28 & Male & Loss or injury of limbs, loss of right leg, needs prosthetics\\ \hline
    P6 &  31 & Female & Cerebral palsy, shaking hands and hard to control hand movements\\ \hline
    P7 &  21 & Male & Cerebral palsy, shaking hands and hard to control hand movements\\ \hline
    P8 &  32 & Female & Spinal cord injuries, wheelchair user\\ \hline
    P9 &  42 & Male & Cerebral palsy, shaking hands and hard to control hand movements\\ \hline
    P10 &  34 & Male & Loss or injury of limbs, loss of one leg, needs crutches\\ \hline
    P11 & 35 & Male & Spinal cord injuries, hands have no feeling\\ \hline
    P12 & 26 & Male & Loss or injury of limbs, right hand has no fingers\\ \hline
    P13 &  28 & Male & Cerebral palsy, shaking hands and hard to control hand movements\\ \hline
    P14 &  26 & Female & Loss or injury of limbs, loss of left leg, needs crutches\\ \hline
    P15 &  35 & Male & Loss or injury of limbs, loss of both of arms\\ \hline
    P16 &  24 & Male & Loss or injury of limbs, loss of legs, needs crutches and prosthetics\\ \hline
    P17 &  27 & Female & Loss or injury of limbs, missing fingers on left hand\\ \hline
    \end{tabular}
    
\end{table*}

\subsection{Participants}
We recruited seventeen (N=17) participants \rv{through online contact with a disability organization} for the study. Table~\ref{participant} shows the demographic information. Twelve were males and five were females. Their average age was 29 years old ($SD = 6$). All participants had some forms of motor impairments that affected their use of mobile phones. Ten participants had arm or hand problems. Specifically, eight had the loss or injury of their limbs, six had cerebral palsy who had shaky hands and difficulty controlling their hand movements. The remaining three had spinal cord injuries, two of whom needed to use a wheelchair\rv{,} and one did not have hand sensation. Seven had their legs amputated and needed prosthetics or crutches. 
Some of them had difficulties speaking clearly or fluently due to the influence of cerebral palsy, but it did not affect our user study. None of them used upper-neck gestures to control devices prior to the study. \rv{The participants were compensated for the study.}

\subsection{Tasks}

\rv{Firstly, we studied the instructions on the official websites of iOS and Android \footnote{\url{https://support.apple.com/en-us/guide/iphone/iph75e97af9b/ios}, \url{https://support.apple.com/en-us/guide/iphone/iphfdf164cac/ios}, \url{https://support.apple.com/en-us/guide/iphone/iphca3d8b4e3/ios}, \url{https://support.google.com/android/answer/9079644?hl=en}, \url{https://support.google.com/android/answer/9079646}} to learn about the commands and corresponding gestures designed for today's touchscreen smartphones. Moreover, we drew inspiration from recent work ~\cite{10.1145/3373625.3416987,fan2021eyelidCAM} for the commands that were supported by eyelid gestures designed for people with motor impairments to interact with touchscreen devices. In the end, we identified 26 commands commonly used for smartphone interactions. 
Based on their similarities, we further clustered} these commands \st{were clustered} into three groups. \textit{Group 1} included twelve \textbf{General commands}, which were Single Tap, Double Tap, Flick, Long Press, Scroll Up, Scroll Down, Swipe Left, Swipe Right, Zoom In, Zoom Out Drag, and Rotate. \textit{Group 2} included ten \textbf{App-related commands}, which were Open the Ap\rv{p}, Move to Next Screen, Next Button, Previous Button, Open the Container (a UI component within an app), Next Container, Previous Container, Move to Next Target App, Open Previous App in the Background, Open Next App in the Background. \textit{Group 3} included four \textbf{Physical Button-related commands}, which were Volume Up, Volume Down, and Screenshot. 

The general commands were obtained from mobile phone systems (iOS$\And$Android), and the app-related commands were inspired by a recent study~\cite{10.1145/3373625.3416987}, which proposed commands such as switching between apps, switching between tabs in an app,  and switching between containers in a tab. The four physical button-based commands were \rv{also inspired by the commands supported by iOS and Android and were} related to common button functions, such as turning the volume up \& down, taking screenshots, and locking the screen.

\subsection{Procedure}
\label{sec:procedure}
\begin{figure*}[h!]
  \centering
  \includegraphics[width=1\linewidth]{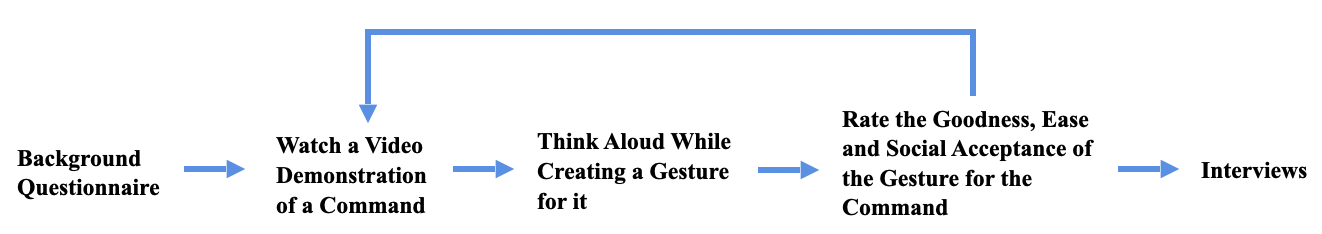}
  \caption{The procedure of the study procedure.}
  \label{fig:procedure}
  \Description{The procedure of the task experiment (i.e., "background questionnaire", "watch a video demonstration of a command", "think aloud while creating a gesture for it", "rate the goodness, ease and social acceptance of the gesture for the command", and "interviews").}
\end{figure*}

\begin{figure*}[htb!]
  \centering
  \subfigure[Zoom In]{
  \begin{minipage}[t]{0.33\linewidth}
  \centering
  \includegraphics[width=2in]{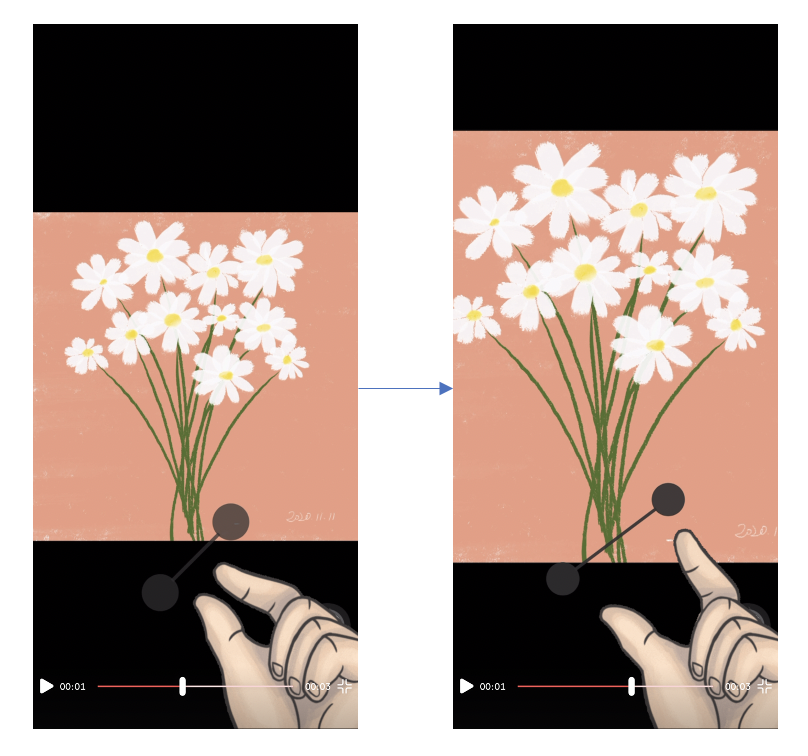}
  \end{minipage}%
  }%
  \subfigure[Next Container]{
  \begin{minipage}[t]{0.33\linewidth}
  \centering
  \includegraphics[width=2.1in]{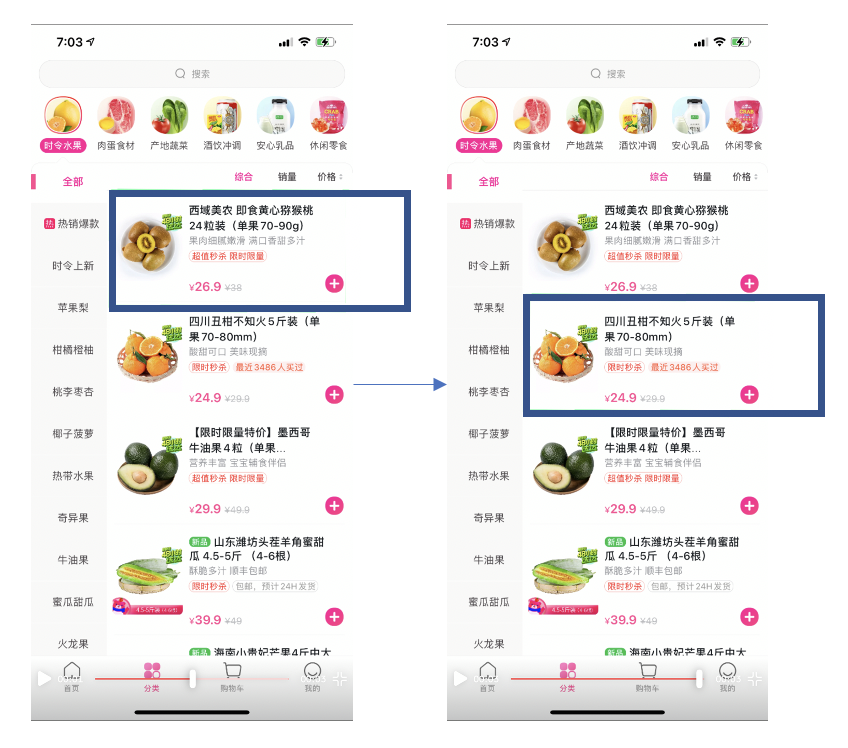}
  \end{minipage}%
  }%
  \subfigure[Volume Up]{
  \begin{minipage}[t]{0.33\linewidth}
  \centering
  \includegraphics[width=2in]{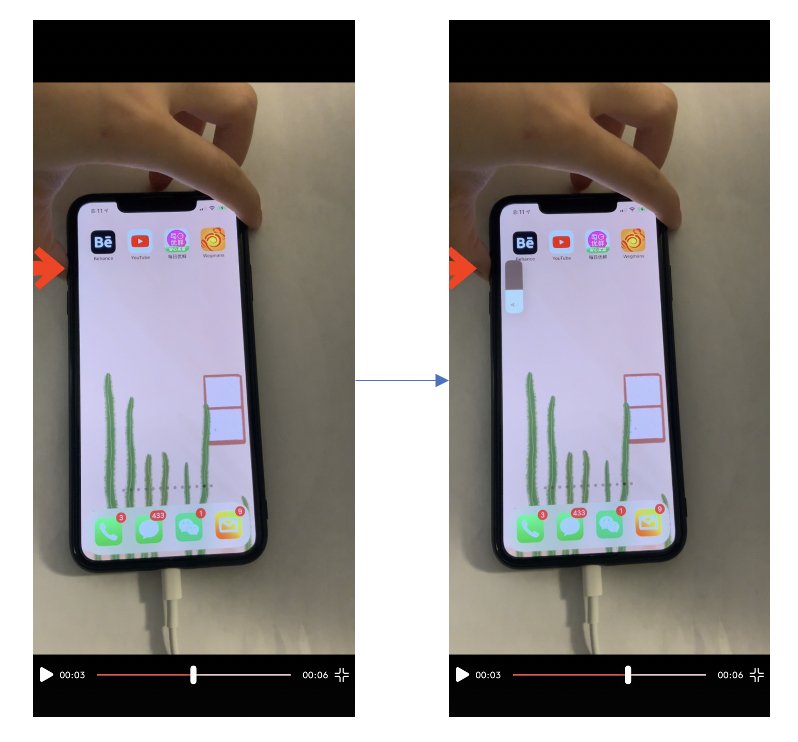}
  \end{minipage}%
  }%
  \centering
  \caption{Video clips: (a) is the video clip for Zoom In, which is an example of enlarging a picture by hand interaction, (b) is the video clip for the Next Container command in the App-related group, showing the target containers that the participants needed to interact with, (c) is the video clip for the Volume Up command in the button-related group, which shows how one of the authors interacted with her mobile phone to increase the volume.}
  \label{fig:video clips}
  \Description{Examples of video screenshots used in the study (i.e., "Zoom In", "Next Container", and "Volume Up").}
\end{figure*}

Figure ~\ref{fig:procedure} shows the study procedure. After answering the background questionnaire, participants were asked to watch a short video clip for a command. We created a short video clip to show each command and its effect on a smartphone. Figure ~\ref{fig:video clips} shows example video clip frames for the command Zoom In, Next Container, and Volume Up. Showing video clips instead of explaining verbally was to ensure the consistent presentation of tasks to participants. After watching the video clip, participants would create an above-the-neck gesture for it and perform the gesture to the moderator. 
To reduce any ordering effects, the task videos for the general commands and physical button-related commands were presented to participants in \st{a }random order. Because there was a logical order among the commands in the app-related group, we kept the order of the tasks in this group to avoid confusion\rv{s}.
During this process, we asked participants to think aloud so that the moderator could better monitor the design process. 
After performing the user-defined gesture, they were asked to rate the goodness, ease\rv{,} and social acceptance of the gestures for the command using 7-point like-scale questions.
Then, participants repeated this process until they created gestures for all commands. 

To reduce the gesture conflicts happening during the study, we asked participants to design different gestures for each command within the same group (i.e., three groups for the general, app-related, and physical\rv{-}button related commands). As for the commands that were not in the same group, we allowed the participants to perform the same gesture. 
However, due to a large number of commands, some participants might forget their previous gestures. Thus, the moderator monitored the gestures already created, and if she found a conflict gesture was proposed, she would remind the participants to change to a different gesture for either the current one or the earlier one that was conflicted with. In addition, participants were also allowed to change their mind\rv{s} if they later wanted to go back and change a previous one. 

All study sessions were conducted remotely through a video conference platform in order to comply with the COVID-19 social distance requirements, and the whole process was video recorded. In total, we collected 442 user-defined above-the-neck gestures (17 participants x 26 commands). 

\subsection{Conceptual Complexity of the Commands}
\label{sec:cc}

\begin{figure}[htb!]
  \centering
  \includegraphics[width=1\linewidth]{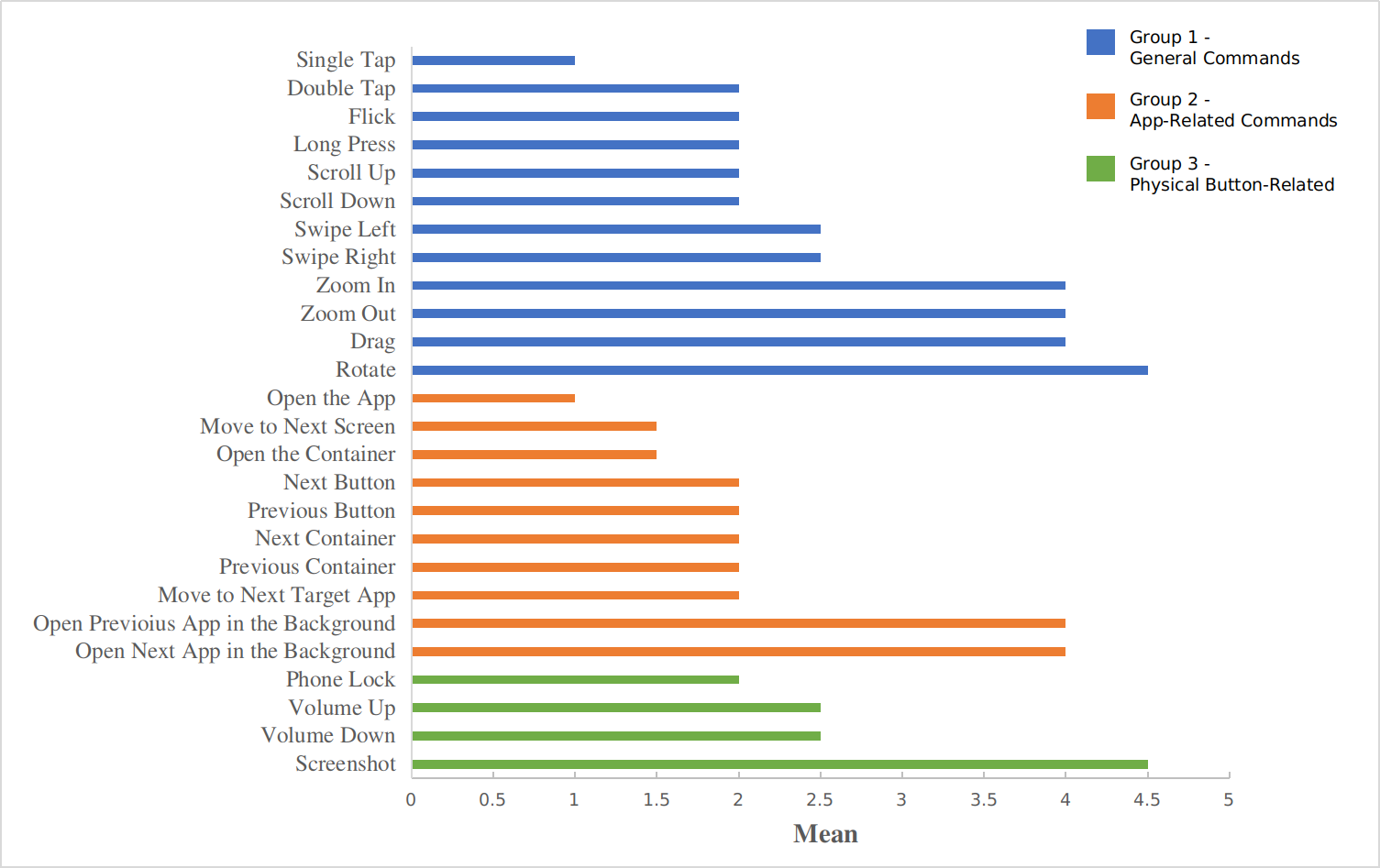}
  \caption{The conceptual complexity of 26 commands, rated by two HCI researchers. The higher the score, the more complex the command \rv{perceived by the HCI researchers}.}
  \label{fig:commands}
  \Description{A bar chart with different colors representing the mean value of conceptual complexity of each command.}
\end{figure}

Before we determined the final user-defined gestures for the commands, we took a step to understand the perceived complexity of the commands. To do so, we calculated the \textbf{conceptual complexity} of each command. 
Conceptual complexity of a command was a concept \rv{widely used in prior works (e.g., Wobbrock et al.'s work~\cite{10.1145/1518701.1518866}, Arefin Shimo et al.'s work~\cite{arefin2016exploring}, and Dingler et al.'s work~\cite{dingler2018designing})}, which measures \rv{the perceived difficulty of the command from HCI researchers' points of view}. 
\st{how difficult to achieve the effect of the command.}
For example, Single Tap \rv{(e.g., tapping a button on the touchscreen)} is a command that \rv{we, as HCI researchers, believed} could be achieved easily with a tap and thus has \st{a }low conceptual complexity. In contrast, Screenshot (e.g., taking a screenshot) requires more \rv{fingers and} steps than a single tap, so it has a relatively higher conceptual complexity. 

To determine the conceptual complexity for each command, two HCI researchers rated the difficulty of completing each command \rv{on a 5-point Likert scale }(1: the relatively easiest, 5: the relatively most difficult) independently \rv{to avoid influencing the other person's point of view}. \rv{A score of 5 means that it was perceived by the researcher as the most difficult among the 26 commands.} \rv{The scores of the two researchers were similar for most commands with only a few that had relatively bigger differences}. Finally, we calculated the average of the scores assigned by the two researchers to get the conceptual complexity of the command. Figure~\ref{fig:commands} shows the conceptual complexity for each command.

\section{Results}
We followed the analysis methods of prior user-defined gesture design papers (e.g., ~\cite{10.1145/1518701.1518866}) to obtained \textit{the gesture taxonomy}, the \textit{gesture agreement score}, the \textit{participants' ratings of the gestures performed}, and the final \textit{user-defined gesture set}. We also analyzed the participants' interviews to understand the design rationales.

\subsection{User-defined Above-the-Neck Gestures Taxonomy}
\begin{table}[htb!]
    \caption{The seven categories of the user-defined above-the-neck gestures, including only eyes, only head, only mouth, eyes and head, eyes and mouth, head and mouth, and eyes, head and mouth, and the gestures within each category}
      \label{Table 1}
    \Description{The table demonstrates the taxonomy of all the gestures, including only eyes, only head, only mouth, eyes and head, eyes and mouth, head and mouth, and Eyes, head and mouth.}
    \begin{tabular}{c|l}
    \hline
     \rowcolor[gray]{0.9}Taxonomy of All Gestures & Breakdown of Each Taxonomy \\
    \hline 
     \multirow{12}{*}{Only Eyes} & blink \\\cline{2-2}
      & gaze \\\cline{2-2}
      & eye-movement \\\cline{2-2}
      & eye size \\\cline{2-2}
      & eyebrows \\\cline{2-2}
      & blink + gaze \\\cline{2-2}
      & gaze + eye size \\\cline{2-2}
      & eye movement + eye size \\\cline{2-2}
      & blink + eye movement \\\cline{2-2}
      & gaze + eye movement \\\cline{2-2}
      & blink + gaze + eye movement \\\cline{2-2}
      & blink + eye size \\\hline
     \multirow{4}{*}{Only Head} & head movement \\\cline{2-2}
      & head distance \\\cline{2-2}
      & head rotation \\\cline{2-2}
      & head distance + head movement \\\hline
     \multirow{5}{*}{Only Mouth} & pout \\\cline{2-2}
     & wide open then close mouth \\\cline{2-2}
     & wry mouth \\\cline{2-2}
     & suck mouth \\\cline{2-2}
     & smile \\\hline
      \multirow{7}{*}{Eyes $\And$ Head} & blink + head movement \\\cline{2-2}
     & eye size + head movement \\\cline{2-2}
     & eye gaze + head movement \\\cline{2-2}
     & eye movement + head movement \\\cline{2-2}
     & gaze + head distance \\\cline{2-2}
     & gaze + head rotation \\\cline{2-2}
     & gaze + eye size + head movement \\\hline
    \multirow{2}{*}{Eyes $\And$ Mouth} & blink + pout \\\cline{2-2}
     & blink + wide open mouth\\\hline
    \multirow{2}{*}{Head $\And$ Mouth} & head movement + pout \\
    \cline{2-2}
     & head movement \\
     & + wide open then close mouth\\
     \hline
     \multirow{2}{*}{Eyes $\And$ Head $\And$ Mouth} & eye movement + head movement \\
     & + wide open mouth\\
    \hline
    \end{tabular}
    
\end{table}

We collected 17×26=442 gestures for all 26 commands and classified them according to different body parts, the eyes, the head, and the mouth. 

\textbf{Gesture Categories}. We grouped the gestures into \textbf{seven categories} based on the body parts involved. These seven categories included \textit{a single body part} and \textit{the combinations of different body parts}: \textbf{only eyes}, \textbf{only head}, \textbf{only mouth}, \textbf{eyes$\And$head}, \textbf{eyes$\And$mouth}, \textbf{head$\And$mouth}, and \textbf{eyes$\And$head$\And$mouth}. For each dimension, we subdivided it according to the gestures that the participants performed. Table~\ref{Table 1} shows the taxonomy of the user-defined above-the-neck gestures.

The \textit{only eyes} category includes five basic types of eye gestures: \textit{blink}, \textit{gaze}, \textit{eye movement}, \textit{eye size}, \textit{eyebrow} and the combination of these basic eye gestures. Blinks include single and double blinks, as well as different numbers of blinks. Gaze is the eyes on the screen, which may differ in the length of time. In addition to moving the eyes up and down, left and right, eye rotating and cross-eye are also considered eye movement. The wide opening, closing, and squinting actions are regarded as the scope of eye size. In addition, we counted eyebrow movements as eye scope, such as having participants squeeze their eyebrows. After combining two or more of these single eye movements, there will be dozens of combinations in total. We got rid of the ones that the participants did not perform and ended up with seven combinations of eye gestures.

The \textit{only head} category includes \textit{head movement}, \textit{head distance}, \textit{head rotation}, and the combination of different head gestures as well. The scope of head movement is turning and tilting the head in different directions. We separated head rotation from the head movement because the amplitude of the rotation is larger and more apparent. In addition, the tilted head included in the head movement is a bit similar to the half-circle rotation, so we thought that distinguishing them could be better understand the participants’ preference for these two different amplitude gestures. Head distance means the distance between the head and the mobile phone screen that people would move heads closer or further from the screen. Among all the combination possibilities of these single head gestures, only the combination of head movement with head distance change was chosen by our participants. 

The \textit{only mouth} category includes \textit{pout}, \textit{open mouth}, \textit{close mouth}, \textit{wry mouth}, \textit{suck mouth}, and \textit{smile}. 

\textit{Combined Gestures}. The above three categories are the gestures involving an individual body part. The participants also made some \textit{combined gestures}, \rv{which were} the gestures with a combination of different body parts. One of the most frequently proposed combined gestures is \textit{the combination of eyes and head gestures}, such as blinking followed by a nod, closing the eyes with the head swinging, gazing with head rotation, etc. The combination of eyes$\And$mouth, head$\And$mouth, had two different varieties in each category. The eyes$\And$head$\And$mouth had only one combination. 

\textbf{Distribution of Gesture Groups}. There were many overlaps between the gestures designed by different participants. After removing overlaps, we found 250 unique user-defined above-the-neck gestures. Among these unique gesture\rv{s}, 44.4\% were only-eye gestures, 14.8\% were only-head gestures, 4.8\% were only-mouth gestures, 30.8\% were eyes$\And$head, 1.2\% were eyes$\And$mouth, 3.2\% were head$\And$mouth, and 0.8\% were eyes$\And$head$\And$mouth. This finding suggests that although participants could use the mouth, the eyes, and the head, they still preferred eye\rv{-}based gestures the most and the head\rv{-}based gestures the second. 

Among all categories, the \textit{Only Eyes} category was most diverse. Moreover, the combinations including eyes (e.g., Eyes \& Head, Eyes \& Mouth) were more common than those performed by other parts (e.g., Head \& Mouth).




\subsection{Determination of the User-defined Gesture Set for the Commands}
To derive the final user-defined gesture set from all gestures proposed by all participants, we collated the gestures included in each command and counted the number of participants performing the same gesture. We resolved conflicts between gestures to obtain the final gesture set. We also calculated the \textbf{agreement score} for each command. \rv{Agreement score was initially proposed by Wobbrock et al.~\cite{wobbrock2005maximizing} and later widely used in studies uncovering user-defined gestures for various platforms (e.g., tabletop, phone, watch, and glasses~\cite{10.1145/1518701.1518866,10.1145/1978942.1978971,dingler2018designing,vatavu2015formalizing,arefin2016exploring}}. \st{It represents the perceived complexity of a command from the perspectives of the end-users (i.e., people with motor impairments).}
\rv{It intuitively characterizes differences in agreement between target users for assigning a gesture to a given command. In general, the higher the agreement score of a command, the better the participants are in agreement \st{for}\rv{with} the gesture assigned to the command.}

\subsubsection{Agreement Score}
We categorized the gestures performed by the participants for each command and then counted how many people made the same gesture. These groups and the number of people in each group were used to calculate the \textbf{agreement score} of the commands. We adopted this method from prior user-defined gesture research~\cite{wobbrock2005maximizing,10.1145/1518701.1518866,10.1145/1978942.1978971} and used the following equation:
\begin{equation}
    A_c = \sum_{P_i
}(\frac{P_i}{P_c
})^2
\end{equation}

In Equation.1, $c$ is one of the commands, $A_c$ represents its \textit{agreement score} based on participants' proposed gestures for this command. The value ranges from 0 to 1.  $P_c$ is the total number of gestures proposed for $c$, which is the number of participants in our case (N=17). $i$ represents a unique gesture. Because different participants proposed \rv{the} same gestures, the number of \textit{unique gestures} was smaller than the total number of proposed gestures. $P_i$ represents the number of participants who propose\st{s} the unique gesture i. 
Take the \textit{Single Tap} command as an example, 17 participants proposed 17 gestures in total, thus $Pc$ equals 17. Among these gestures, there were seven unique gestures. There were 7, 4, 2, 1, 1, 1, and 1 participants who proposed each of the seven unique gestures respectively. As a result\rv{,} the agreement score of \rv{the} Single Tap command was calculated as follows: 
\begin{equation}(\frac{7}{17})^2+(\frac{4}{17})^2+(\frac{2}{17})^2+4
(\frac{1}{17})^2=0.25\end{equation}.


\begin{figure}[htb!]
  \centering
  \includegraphics[width=1\linewidth]{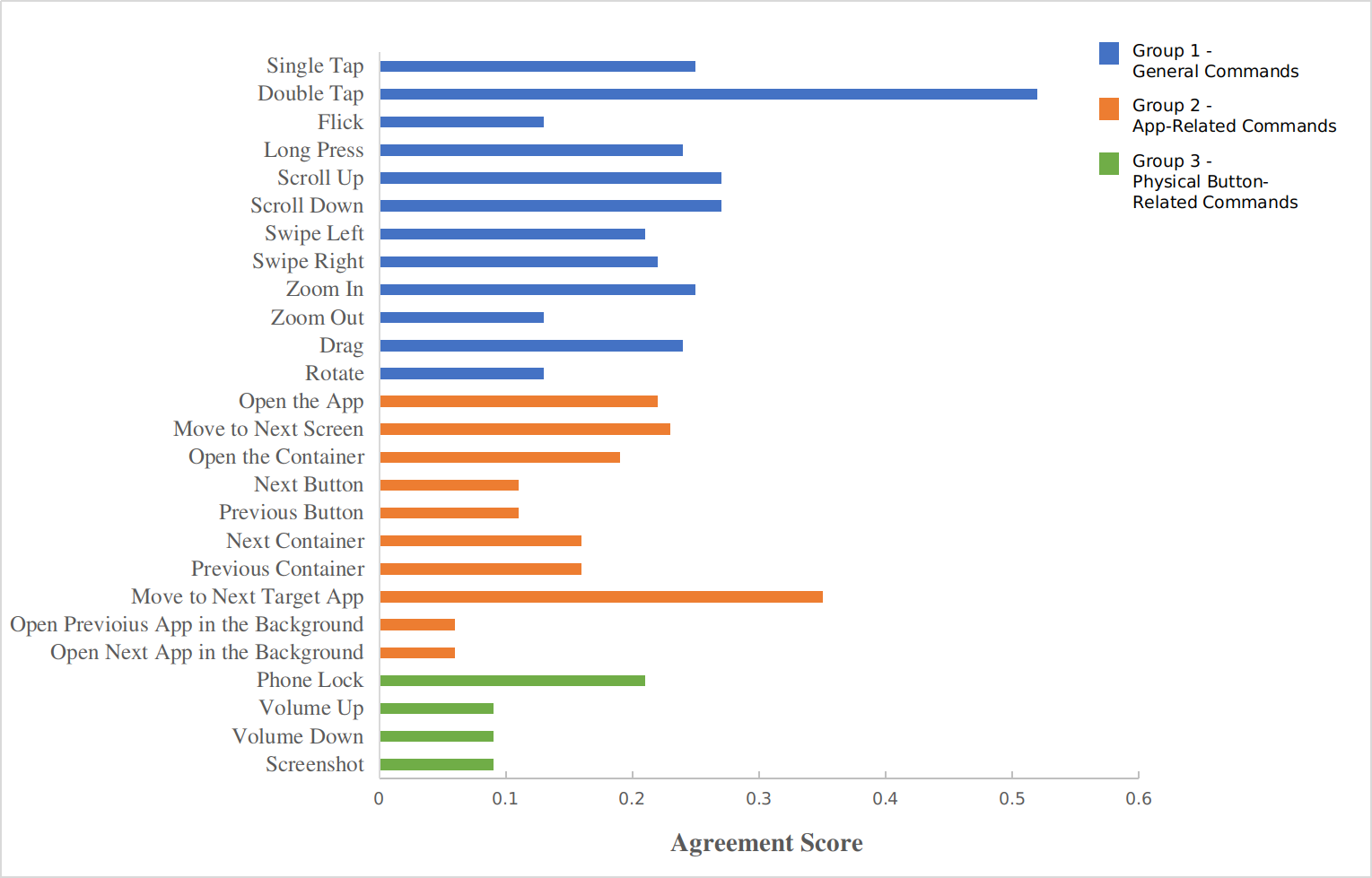}
  \caption{The \textit{agreement scores} of the 26 commands. The higher the score, the higher the participants' consensus on which gesture(s) should be assigned}
  \label{fig:agreement}
  \Description{A bar chart with different colors representing the agreement score of each command.}
\end{figure}

Figure~\ref{fig:agreement} shows the agreement score of the gestures proposed for each command. The commands were arranged in the same order as the conceptual complexity as in Figure~\ref{fig:commands}. In general, the higher the agreement score, the higher the participants' consensus on which gesture(s) should be assigned. The agreement score of \textit{Double Tap} was high, which indicated that participants more agreed on which gesture(s) should be allocated to this command. In contrast, the agreement score of \textit{Rotate} was relatively lower, which indicated that participants proposed more diverse gestures for it and less agreed on which gesture should be allowed to it. 


\subsubsection{Conceptual Complexity vs. Agreement Score}

As we explained in Sec.\ref{sec:cc}, the \textit{conceptual complexity} is a measurement of the perceived complexity of commands from \textbf{researchers' perspectives}. In contrast, \textit{agreement score} is a measure of perceived complexity of commands from \textbf{end-user's perspectives}. If the perception of researchers aligns with the perception of end-users (i.e., people with motor impairments), then we should expect to see a correlation between the two measures.
We did Pearson Correlation Test and \st{interesting, we}found no significant correlation between the agreement score and conceptual complexity score of each command (r=-.38, p=.05). In other words, the commands given a low conceptual complexity score by the researchers\st{,} did not result in a high agreement score. This suggests that there was a discrepancy between the researchers' understanding of the complexity of the commands and gestures and that of the target users'\rv{. This finding} further highlights the necessity to involve end-users into the design process to design user-defined gestures that they\rv{, instead of researchers,} would perceive easy-to-use. 


\subsubsection{Gesture Conflict}
We found that in some cases participants proposed different gestures for a command. Thus, we needed to resolve the conflicts to assign one gesture for each command. 

\begin{figure*}[h]
  \centering
  \includegraphics[width=1\linewidth]{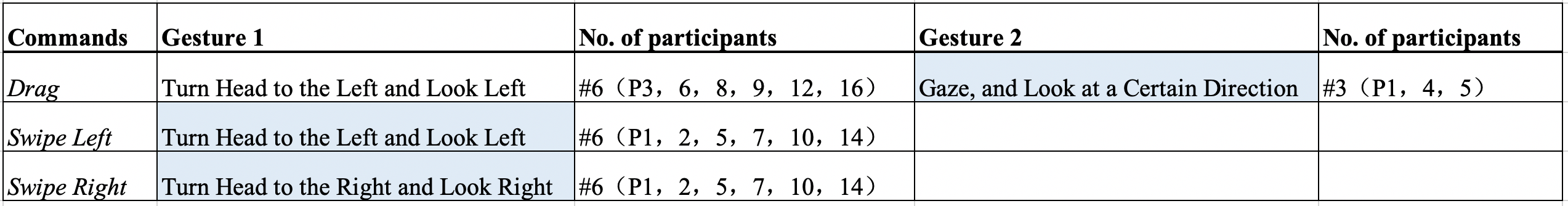}
  \caption{Gesture conflict 1: a single command (e.g., Drag) conflicted with paired commands (e.g., Swipe Left$\And$Swipe Right). In this case, we assigned gesture 2 to Drag and gesture 1 to Swipe Left$\And$Swipe Right. The column of No.of participants means how many participants performed the same gesture, including the participant number. The gestures were arranged in the descending order of No.of participants.}
  \label{fig:conflict 1}
  \Description{A table shows how we solved the gesture conflict when a single command conflicted with paired commands.}
\end{figure*}

Our conflict resolution strategy was as follows. When the same gesture was allocated to both a single command (e.g., Drag) and a paired commands (e.g., Swipe Left$\And$Swipe Right), we would prioritize the gesture to the paired ones because the cost for finding alternative gestures for the paired commands was higher than that for a single command. 
After allocating the conflicted gesture to the paired commands, we allocated the gesture proposed by the second-highest number of participants for this single command to it. 
Figure~\ref{fig:conflict 1} illustrates the process with an example.
The same gesture ``Turn Head to the Left and Look Left'' was proposed by the same number of participants (N=6) for both the single command (i.e., Drag) and the paired commands (i.e., Swipe Left$\And$Swipe Right). Our resolving strategy would assign this gesture to the Swipe Left command. Next, we allocated the gesture that was proposed by the second-highest number of participants to the Drag command. In this case, it was ``Gaze, and Look At a Certain Direction.''

\begin{table}[htb!]
    \caption{The final user-defined gestures for all commands. Most commands have one allocated gesture (i.e., Gesture 1) and a few have two or three gestures (i.e., Gesture 2 and Gesture 3)}
      \label{gesture set}
    \Description{The table demonstrates our final gesture set. For each of the 26 commands, we assigned the most appropriate gesture or gestures to it.}
    \begin{tabular}{p{0.2\textwidth} | p{0.3\textwidth}|p{0.2\textwidth} | p{0.2\textwidth}}
    \hline
     \rowcolor[gray]{0.9}Commands & Gesture 1 & Gesture 2 & Gesture 3 \\
    \hline 
    Single Tap & eyes blink once & &  \\ \hline
    Open the App & eyes blink once & &\\ \hline
    Move to Next Screen & turn head to the left and look left & & \\ \hline
    Open the Container & eyes blink once & & \\ \hline
    Double Tap & eyes blink twice &  & \\ \hline
    Flick & raise head & & \\ \hline
    Long Press & gaze for 3-5s & & \\ \hline
    Next Button & look right, then blink once & & \\ \hline
    Previous Button & look left, then blink once & & \\ \hline
    Next Container & look downward & & \\ \hline
    Previous Container & look upward & & \\ \hline
    Move to Next Target App & look right & & \\ \hline
    Scroll Up & raise head and look upward & look upward & \\ \hline
    Scroll Down & lower head and look downward & look downward & \\ \hline
    Phone Lock & close eyes for 3s & & \\ \hline
    Swipe Left & turn head to the left and look left & & \\ \hline
    Swipe Right & turn head to the right and look right & & \\ \hline
    Volume Up & blink right eye & & \\ \hline
    Volume Down & blink left eye & & \\ \hline
    Zoom In & wide open eyes & & \\ \hline
    Zoom Out & squint eyes & & \\ \hline
    Drag & gaze, and look at a certain direction & & \\ \hline
    Open Previous App in the Background & raise head, then turn head to the right, then blink eyes & & \\ \hline
    Open Next App in the Background & raise head, then turn head to the left, then blink eyes & & \\ \hline
    Screenshot & eyes blink three times & & \\ \hline
    Rotate & turn head to the left, and look at the screen & eyes look counter-clockwise & tilt head \\ \hline
    \end{tabular}
    
\end{table}

\subsubsection{Final User-defined Above-the-Neck Gesture Set}
Table~\ref{gesture set} shows the final gestures for each command. Most of the commands have only one allocated user-defined gesture (i.e., Gesture 1 in the table). However, there are three commands that had more than one gesture allocated. This is because these commands have more than one gesture \st{that were }proposed by the same number of participants. 
As shown in Table~\ref{gesture set}, twenty-three commands were assigned 1 gesture, two commands had 2 gestures, and one command had 3 gestures.

\begin{figure*}[htb!]
  \centering
  \includegraphics[width=1\linewidth]{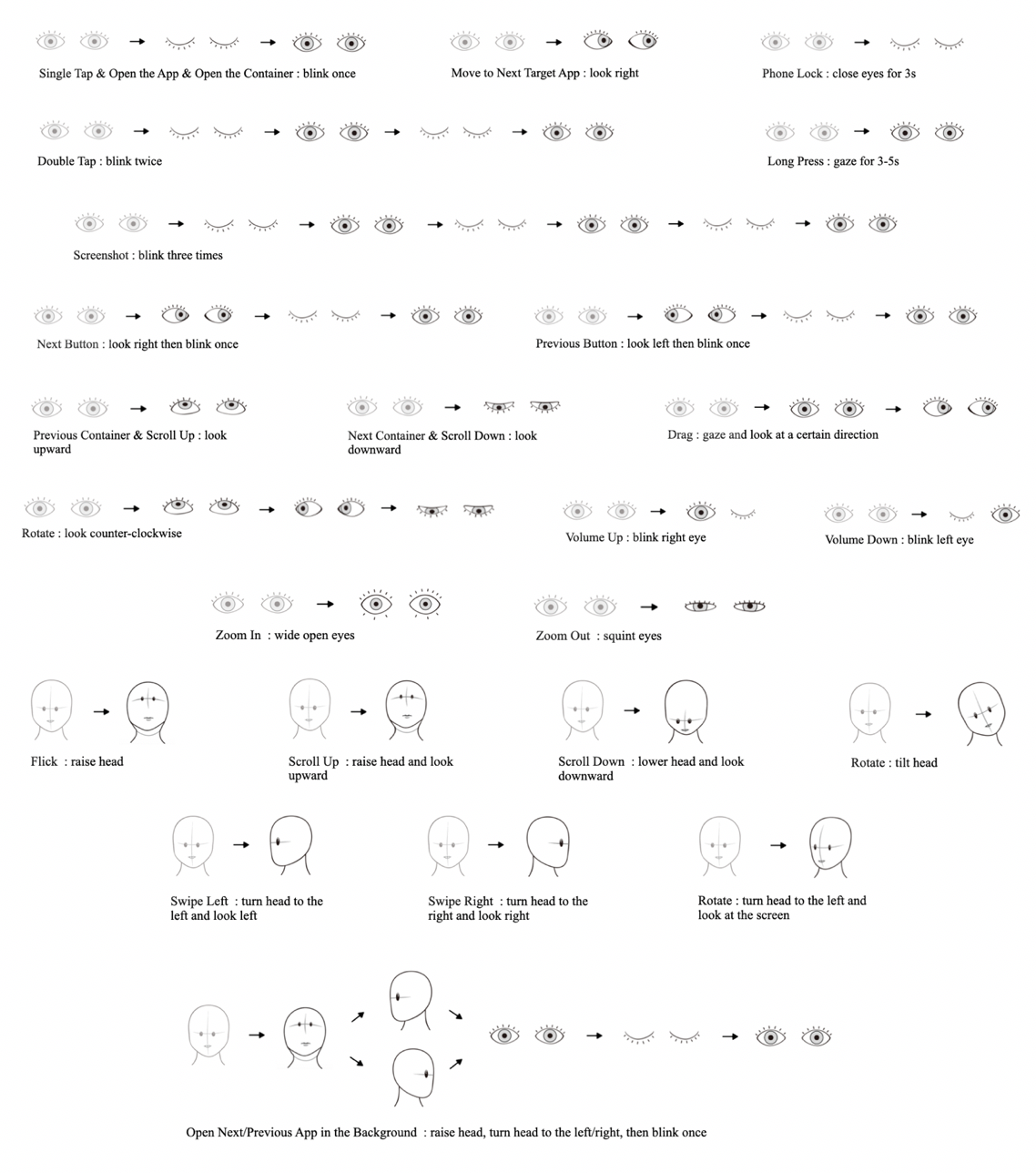}
  \caption{Visual illustrations of the final user-defined above-the-neck gestures for the 26 commands.}
  \label{fig:sketch}
  \Description{The sketch of our final gesture set.}
\end{figure*}

Figure~\ref{fig:sketch} further illustrates the final user-defined above-the-neck gestures for the 26 commands.
There were 30 gestures in the final gesture set. 20 were only-eyes gestures, 3 were only-head gestures, and 7 were eyes$\And$head gestures.


We found continuity in our final gesture set. When two commands were related, such as single tap and double tap, the final assigned gestures were blinking once and blinking twice, which were very strongly correlated. There is also a strong symmetry. For example, for swipe left and swipe right, the gestures were looking at left and right and turning head to the left and right; and for zoom in and zoom out, the gestures were wide-open eyes and squint eyes. These gestures were also logical, such as the phone lock was closing eyes.

\subsubsection{Subjective Ratings of the Gestures}
We asked participants to rate the goodness, ease of use, and social acceptance of each gesture they performed. We divided the gestures under each command into large groups and small groups. The large group was defined as the number of people making that gesture over the number of people making the other gestures, and it was usually the gesture that was selected in the gesture set. The rest are the small groups. We compared the participants’ average goodness (mean of large groups=5.94, mean of small groups=5.94), ease of use (mean of large groups=5.80, mean of small groups=5.72), and social acceptance (mean of large groups=5.69, mean of small groups=5.70) of the gestures in large groups with those in small groups. There was no significant difference between them.  We also did Pearson Correlation Test and found that the conceptual complexity of the commands did not have a strong correlation of their goodness (r=-.42, p=.04), ease of use (r=-.20, p=.34), or social acceptance (r=-.40, p=.05). The reason was many participants commented while doing the test that "I choose to make this gesture because I think it is good," resulting in a high rating score, which proved why a gesture was performed by fewer participants and command had a high conceptual complexity score, but the rating was not low. 


\subsection{Perceptions of the User-defined Gestures}
We present the following insights learned from participants' feedback about user-defined gestures.

\subsubsection{Easy to Use and Understand}
After participants made a gesture, we asked them why they chose that gesture, the most common feedback was that it was simple to understand, easy to do, and understandable.
P13 explained, \textit{``Because I feel that these gestures are good, simple, easy to perform, so I choose them.''}, and P16 argued, \textit{``It is easy to understand and belongs to the normal range of head movement.''}

In addition to the gesture itself was simple to do, but also with the previous use of technical devices. Many participants thought about creating gestures to complete commands in conjunction with how they usually used their mobile phones with their hands. For example, when I asked them to swipe to the next page, most of them turn their heads to the left because they also swiped to the left when operated by hand. P11 used his computer more often than his phone, so he considered his habits of using the computer when creating gestures. For example, when doing the single tap command, he blinked his left eye and said, \textit{``Operation is similar to clicking with the left mouse button''-P11}. Using different mobile phone models also affected the gestures. Many Android phone users chose to blink three times when taking a screenshot because they usually used three fingers to pull down for taking a screenshot with their Android phones, so this gesture was better for them to understand, but it might not be for iPhone users. 

\subsubsection{Memorability}
During the testing process, many participants gave feedback that they forgot what gestures they had done before. When P14 was doing command 15, she said she couldn't remember what gestures she had done. In addition to not remembering, some participants deliberately chose to make the gestures they could easily remember. 

\subsubsection{Duration of Gestures}
There were some gaze or close-eyes gestures, and these gestures would need to take into account the issue of how long the eyes were gazed at or closed. Although some participants mentioned 5 seconds or 10 seconds, most were 2 seconds or 3 seconds. The time length should be moderate, and as long as the differences could be distinguished from the length of time, it would not have to take too long. \textit{``Would 2 seconds or 3 seconds be better? It is clearer to stop for 3 seconds.``-P8}; \textit{``It would be inconvenient if closing the eyes for too long.''-P12} 

\subsubsection{Recognition}
Many participants considered whether the action would be too subtle for the phone to recognize when they did it. For example, when considering the gaze time, the participants thought about whether the time was too short for the phone to recognize. Many participants preferred head gestures because they thought eye movements were too small to be identified. They also considered whether the gesture was confused with our natural movements or random small movements that a spontaneous movement would trigger a command. Thus, some participants made a distinction between the gestures they performed and what we usually did spontaneously. \textit{``I feel that the head forward may not be very sensitive to identify.''-P14}; \textit{``An occasional small action may affect the phone recognition.''-P11}

\subsubsection{Self-condition}
The participants were people with motor impairments, and their motor impairment problems could cause some eye or facial movement difficulties, especially eye closure difficulties. Some participants had difficulty closing a single eye, so they chose to make the gestures in both eyes together. Some participants had difficulty closing the left or right eye, and they chose to make the gesture with the eye that had no difficulty closing. P3 was left-handed. He was more willing to use his left eye to make the gesture. \textit{``I can close both eyes, but cannot close only one eye, so I do it with both eyes.''-P13}

\subsubsection{Social Acceptance}
In the comments and observations, we found that these participants were very concerned about the eyes of others. They would consider whether the gesture would be too exaggerated and did not want to attract attention. They would like to choose some simple gestures so that they would not look strange. \textit{``Doing eye gestures will still take into account the feelings of others.''-P3}; \textit{``Simple, does not attract special attention or disturb others.''-P11};
\textit{``I don't want people to look at me differently.''-P14}
\section{Survey Study}
We identified user-defined above-the-neck gestures by resolving conflicts in the original gestures created by a group of 17 people with motor impairments as illustrated in the previous section.  One follow-up question would be: \textbf{\st{what are the general user acceptance of these user-defined gestures} what are the more appropriate gestures among these user-defined ones for the users who would use \rv{the} gestures to interact with smartphones}? \rv{Our gesture elicitation study was to create a gesture set for people with motor impairments to interact with the touchscreen smartphones without touch. However, able-bodied people might also encounter the same difficulties in many scenarios and find such gestures useful. For example, it would be hard to operate the phone by hand when people's hands are occupied, for example, while carrying bags or with wet or dirty hands. As a result, we included people without motor impairments in this survey study as well to understand their preferences for the user-defined gestures designed by people with motor impairments.}
To answer this question, we conducted an online survey to validate the \st{acceptance and} agreement of the user-defined gesture set by people with and without motor impairments.

\subsection{Method}
\begin{figure}[htb!]
  \centering
  \includegraphics[width=0.5\linewidth]{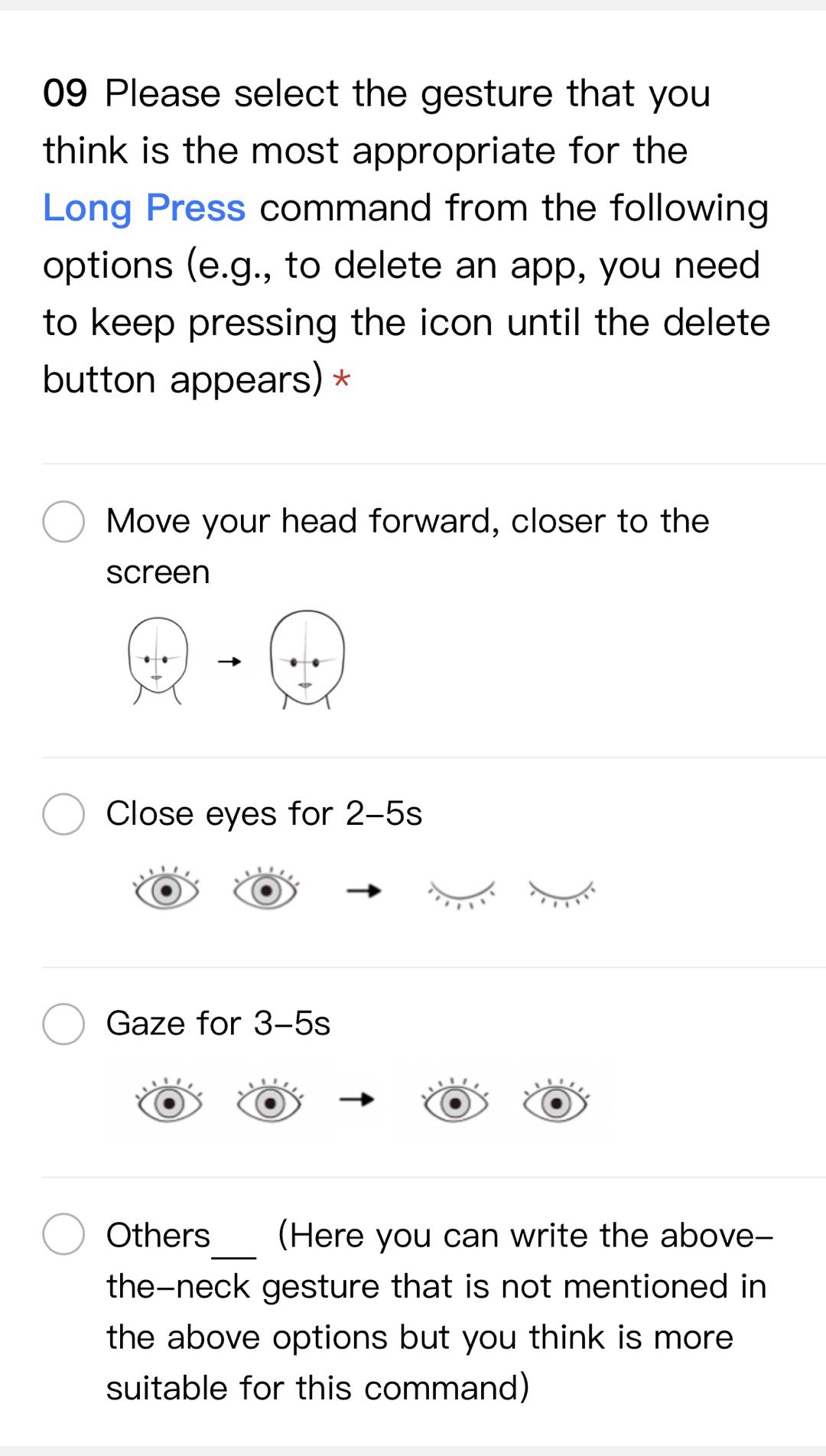}
  \caption{Survey example: the screenshot of the Long Press question.}
  \label{fig:question1}
  \Description{The screenshot of one question in the survey.}
\end{figure}
In the survey, participants were first asked questions about their physical condition and some general demographic questions. Then, we asked about each of the 26 commands proposed in the previous user study. In each question, we asked participants to choose the most appropriate gesture for that command from three candidate options. These candidate options were derived from previous studies and were the gestures that participants most frequently performed for each command. These options were arranged in random order to reduce potential order bias. We also provided an additional option for participants to write out what they thought was most appropriate if they felt none of the three options were suitable.

Figure~\ref{fig:question1} shows the screenshot of the question for the Long Press command in the survey. The survey was published in an online format and took an average of 5 minutes to complete.


\subsection{Participants}
Twenty-four participants volunteered in the survey study, including 10 people with motor impairments and 14 people without motor impairments. \rv{Ideally, it would be best to recruit people with motor impairments who do not have any prior knowledge about the study. However, due to the difficulty in recruiting people with motor impairments during the pandemic, we recruited the 10 participants with motor impairments who took the previous gesture elicitation study. To mitigate the potential memory effect, we did intentionally conduct the survey study one month after the gesture elicitation study. With a one-month time gap between the two studies, it was unlikely that these participants would still remember the commands or how they assigned the gestures to the commands a month ago. For the 14 participants without motor impairments, we recruited them through both online and offline means.} The average age of all participants was 29 years ($SD=3$). 


\subsection{Results}
\begin{figure*}[htb!]
  \centering
  \includegraphics[width=1\linewidth]{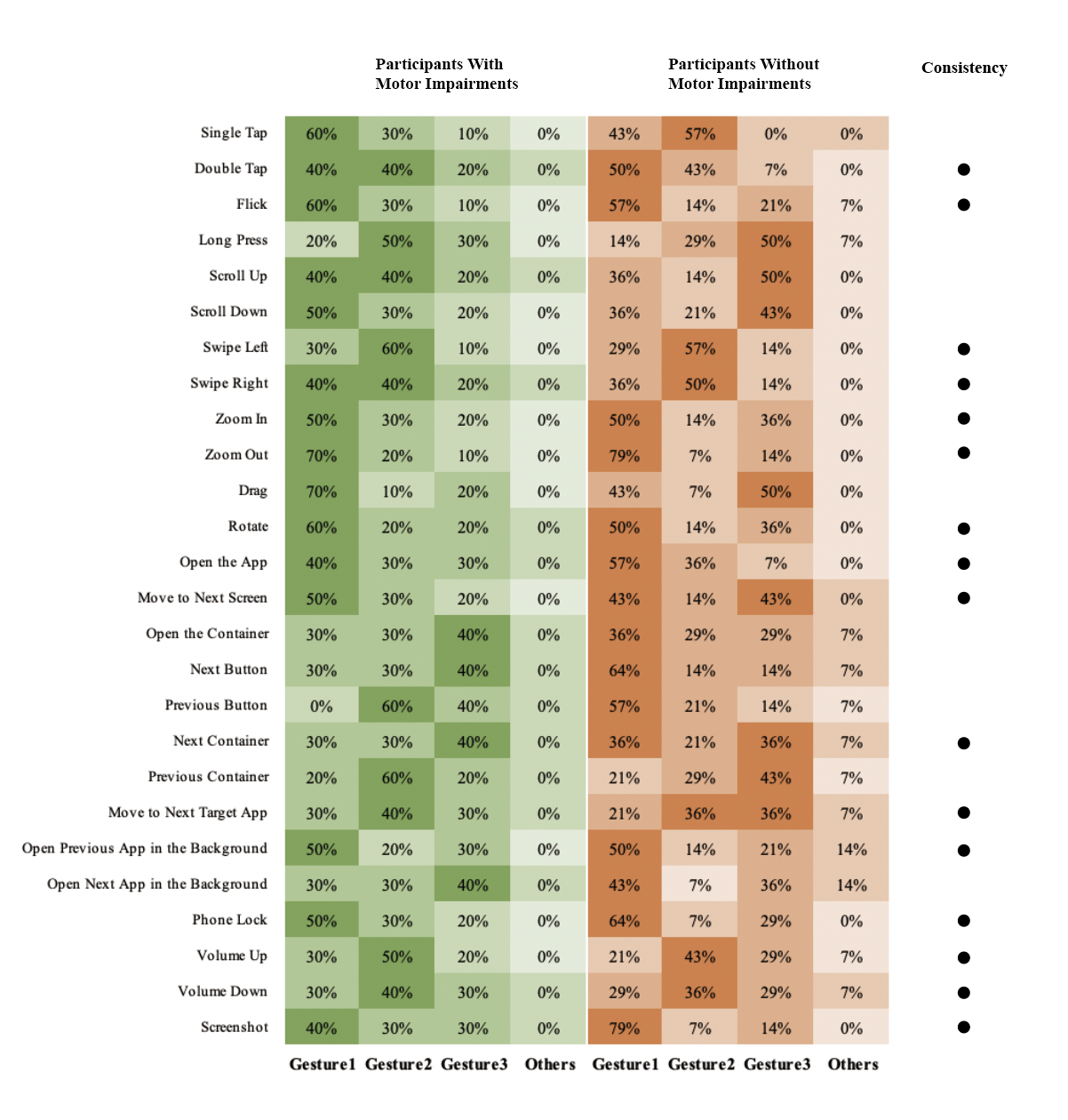}
  \caption{Survey results: participants with motor impairments in green color, participants without motor impairments in orange color. The darkest green/orange color represents the gestures with the highest number of selections for that command. The lightest green/orange color represents the gesture with the lowest number of selections.  \rv{The black dots in the consistency column represent the two groups of people making the same choices.}}
  \label{fig:survey}
  \Description{A heat-map of the survey results in different colors, representing the results of all participants, participants with motor impairments and participants without motor impairments.}
\end{figure*}

In Figure~\ref{fig:survey}, we showed \st{the results of the responses of 24 survey participants in blue color,}
the results of the participants with motor impairments in green color and the results of the participants without motor impairments in orange color. As we mentioned that there were three most frequently performed gesture options for each question, and they were all performed by the participants in the previous user study. 
These three options in all questions were collectively referred to as \textit{Gesture 1}, \textit{Gesture 2}, and \textit{Gesture 3}, in descending order of the number of times participants performed them in the user study. Of the results for all participants in the survey study, \textit {Gesture 1} was the first choice for 17 out of 26 commands and the second choice for the other 6 commands. From the results of the ten motor-impaired individuals, \textit{Gesture1} was preferred in 15 commands and the second choice for 9 commands. From the results of fourteen non-motor-impaired individuals, \textit{Gesture1} was the first choice for 15 commands and the second choice for 8 commands. 
\rv{Furthermore, Figure~\ref{fig:survey} also includes a "consistency" column to indicate the commands for which both participants without and with motor impairments agreed on the most popular gesture options. Out of the 26 commands, both user groups agreed on 16 commands (62\%), which} 
\st{The overall results} indicate\rv{s} that our gesture set had a reasonably high \st{acceptance and }agreement between participants with and without motor impairments.

The agreement of \textit{Gesture 1} by the motor-impaired population was most on general commands, and the non-motor-impaired population was most on app-related and button commands. A possible reason for this was that as the complexity of the command increased, the gestures became more complex, and the combination of different above-the-neck parts and gestures involved increased. Some app-related commands were more complicated than general commands, and people were more likely to create complex gestures. For participants with motor impairments, in addition to considering the appropriateness of the gestures, it was also essential to consider their physical conditions when creating complex gestures, making it more difficult to reach an agreement.

We also noticed that none of the people with motor impairments chose \textit{Others} options in the survey. The possible reason for this was that they had already done the earlier user study. Therefore, the three options provided were either the same or similar to what they had demonstrated in the user study, which led to their agreement on these three options. While in the survey for non-motor-impaired participants, we found that one participant chose the option \textit{Others} very frequently. We checked her responses to see that she indicated that she did not come up with more appropriate gestures but disagreed with our provided options. We then interviewed her briefly, and she told us, \textit{``some gestures are so complicated that no one would find it easier than using their hands unless they had paraplegia''}. Another participant also chose \textit{Others} options for two of the commands, but did not suggest any gestures that she would prefer to use.

Through the survey, we found that the choices of gestures by people with and without motor impairment were mostly consistent with our gesture set. However, there were also some differences due to physical impairments or lack of understanding of people with motor impairments by people without motor impairments. In general, our gesture set is universally applicable to people with and without motor impairments and is accepted and recognized by the public.

\section{Discussion}
We discuss the implications of the user-defined above-the-neck gestures design and users with motor impairments.

\subsection{Key Takeaways}
By involving people with motor impairments \st{into}\rv{in} the design process and \st{revolving}\rv{resolving} conflicts in the proposed gestures, we uncovered a set of user-defined above-the-neck gestures and how people with motor impairments would want to use these gestures to execute commands on a touchscreen mobile device. 
All of the participants of the user study mentioned that they would like to use these gestures to interact with mobile phones in the future. 
Moreover, we learned that people with motor impairments preferred gestures that were simple, easy to remember, and had high social acceptance. Although they had the freedom to include eyes, mouth, and head into the design of the gestures, gestures involving eyes were still the most diverse and preferred, followed by the gestures that combined eyes and head. This finding is consistent with the prior works~\cite{kyto2018pinpointing, sidenmark2019eye} that people like to add head movements to eye-based gestures.  
Our survey study with people with and without motor impairments found that these user-defined gestures were generally agreed by people with and without motor impairments.

\subsection{Design Considerations for User-Defined Gestures}
Compared to the eyelid gestures designed by the researchers~\cite{10.1145/3373625.3416987,fan2021eyelidCAM}, our user-defined gestures are unique in two aspects. First, our gestures are grounded in the preferences and creativity of people with motor impairments. Second, our gestures are more diverse, which not only include\st{d} eyelids but also eye motion and other body parts (e.g., head and mouth), and are more expressive and can be used to accomplish more commands. However, one must be wary of the downside of a more diverse set of user-defined gestures.

First, as the number of user-defined gestures increases, the efforts for remembering the mapping between the user-defined gestures and the corresponding commands also increases. Indeed, some of the user study participants asked us in an apprehensive tone \st{that }whether they would have to remember all the gestures they created throughout the study. Thus, it is worth investigating how best to help people with motor impairments make use of these user-defined gestures with \rv{the }minimum burden of memorization. One possible solution might be to suggest relevant user-defined gestures based on the initial input of the user. For example, if the user starts to close one eye, then the system could recommend a much smaller set of gestures that start with "close one eye." 

Second, some participants were concerned about the long\rv{-}term or frequent use of some gestures, which might develop a bad habit. For example, P15 designed a gesture that required him to tilt his mouth to the left side to use it, and he was worried about creating a bad habit of tilting the mouth. Perhaps when designing user-defined gestures for people with motor impairments, we need to consider not only the gestures' simplicity and social acceptance but also the gestures' long-term health implications. 

Third, we derived a standard set of user-defined gestures by resolving conflicts between the gestures proposed by different participants. While the standard set might be the most applicable set for a group of people with motor impairments, the set might not be optimal for a particular user. \rv{Users may have different physical conditions and habitual perceptions.} For example, if a user could not close her left eye well, then she should be able to skip all the gestures involving closing the left eye and define her alternatives. \rv{If the users prefer to use their eyes, they should have the flexibility to use only-eye gestures instead of head-involved gestures. In addition, as mentioned above, if users are concerned about potential negative effects of performing the same gesture too often on their facial expressions, they could choose to allocate multiple gestures to trigger one command.} Thus, it is important to make this standard set of user-defined gestures applicable so that an individual user with motor impairment could customize it. \rv{The personalization could be users assigning multiple gestures to one command or assigning the same gesture to different commands.}  

Lastly, social acceptance was a key factor that people with motor impairments considered when designing the gestures. However, little is known whether people with motor impairments perceive as socially unacceptable are really unacceptable from the public's perspective and vice versa. Furthermore, it also remains unclear what gestures are more socially acceptable. Perhaps gestures that are small in amplitude and consistent with normal daily routine activities might be more \st{more }socially acceptable. One interesting question would be \rv{to} understand the relative social acceptability of our user-defined gestures.

\section{Limitations and Future Work}

\rv{\textbf{Potential Effect of The Presentation Order of Commands}.}
\rv{When defining gestures, the participants were given the commands randomly. Thus, they were not able to know in advance the complexity of all commands or whether there were similar or symmetrical commands. 
Our approach to alleviating this potential presentation order effect, as stated in Section~\ref{sec:procedure}, was allowing the participants to change the gestures assigned earlier anytime during the study. If they had difficulty keeping track of the assigned gestures, for example, forgetting about what gestures had been already assigned, they could ask the moderator to remind them. Our approach was promising based on the final gesture set shown in Table~\ref{gesture set}. For example, different commands had been assigned with different gestures, and symmetrical gestures were given to symmetrical commands. Nevertheless, it is interesting to investigate whether participants would adopt different strategies to allocate gestures to commands if they were allowed to know all the commands upfront.}


\rv{\textbf{Command Selection}.} We explored 26 commands commonly used on a touchscreen smartphone. However, these commands are not exhaustive. For example, participants mentioned \st{about} other commands, such as returning to the main screen, unlocking the phone, \rv{and back to the previous page}. \rv{In addition, our classification of commands was based on the prior work ~\cite{10.1145/3422852.3423479,10.1145/3373625.3416987,fan2021eyelidCAM}, and we divided the 26 commands into three categories based on the task goals. Different classification methods may also affect the participants' choice of gestures.} Future work could apply the same principles to define user-defined gestures for additional commands. However, one challenge is to resolve conflicts among the gestures allocated for different gestures, which has proved to be increasingly challenging with more commands to cater.

\rv{\textbf{Common vs. Personal User-defined Gestures}. Our study aimed to identify a set of common user-defined gestures for the commands often performed on a touchscreen smartphone based on the feedback from a group of people with motor impairments. Thus, we believe that this set of common user-defined gestures is a good starting point for people with motor impairments to interact with smartphones without touch. However, we acknowledge that people with motor impairments have different residual motor abilities and may be able to or prefer to perform different gestures. Thus, it is imperative to investigate how best to allow people with motor impairments to design a \textit{personal} set of user-defined gestures tailed to their specific motor abilities, such as a recent work by Ahmetovic et al.~\cite{10.1145/3447526.3472044}.} 

\rv{\textbf{Differences Between People With and Without Motor Impairments.}} Our survey study revealed the preferences of people with and without motor impairments. However, it remains unknown why people without motor impairments preferred the same or different user-defined gestures for the same command or whether they care about social acceptance of these gestures as much as people with motor impairments do. 


\rv{\textbf{Potential Effects of Age and Culture.}} Our participants were primarily young and middle-aged people. It remains an open question of whether age plays a role in user-defined gestures. Future work could replicate the study with older adults with motor impairments and examine whether the user-defined gestures are applicable across different age groups and whether there are specific user-defined gestures that are more preferred by an age group. 

\rv{Our participants lived in Asia, and they were used to the culture of the East. The creation and preference of user-defined gestures might be affected by culture, and the social acceptance of the gestures was also likely related to the social norms and cultures they lived in. Future work could explore user-defined gestures for people with motor impairments in different cultures and compare the similarities and differences to understand cross-culture and culture-specific user-defined gestures.}
\section{Conclusion}
We have adopted a user-centered approach by involving people with motor impairments to design user-defined above-the-neck gestures for them to interact with mobile phones. By analyzing the 442 gestures and resolving conflicts, we have arrived at a set of user-defined gestures. The participants were excited about the convenience the gestures could bring to them. They preferred gestures that were simple, easy to remember, and had high social acceptance. 
Our follow-up survey study results found that the user-defined gestures were well received by both people with and without motor impairments. Finally, we also highlight the design considerations and future work.





\bibliographystyle{ACM-Reference-Format}
\bibliography{main.bib}


\begin{thebibliography}{38}


\ifx \showCODEN    \undefined \def \showCODEN     #1{\unskip}     \fi
\ifx \showDOI      \undefined \def \showDOI       #1{#1}\fi
\ifx \showISBNx    \undefined \def \showISBNx     #1{\unskip}     \fi
\ifx \showISBNxiii \undefined \def \showISBNxiii  #1{\unskip}     \fi
\ifx \showISSN     \undefined \def \showISSN      #1{\unskip}     \fi
\ifx \showLCCN     \undefined \def \showLCCN      #1{\unskip}     \fi
\ifx \shownote     \undefined \def \shownote      #1{#1}          \fi
\ifx \showarticletitle \undefined \def \showarticletitle #1{#1}   \fi
\ifx \showURL      \undefined \def \showURL       {\relax}        \fi
\providecommand\bibfield[2]{#2}
\providecommand\bibinfo[2]{#2}
\providecommand\natexlab[1]{#1}
\providecommand\showeprint[2][]{arXiv:#2}

\bibitem[\protect\citeauthoryear{Ahmetovic, Riboli, Bernareggi, and
  Mascetti}{Ahmetovic et~al\mbox{.}}{2021}]%
        {10.1145/3447526.3472044}
\bibfield{author}{\bibinfo{person}{Dragan Ahmetovic}, \bibinfo{person}{Daniele
  Riboli}, \bibinfo{person}{Cristian Bernareggi}, {and} \bibinfo{person}{Sergio
  Mascetti}.} \bibinfo{year}{2021}\natexlab{}.
\newblock \bibinfo{booktitle}{\emph{RePlay: Touchscreen Interaction
  Substitution Method for Accessible Gaming}}.
\newblock \bibinfo{publisher}{Association for Computing Machinery},
  \bibinfo{address}{New York, NY, USA}.
\newblock
\showISBNx{9781450383288}
\urldef\tempurl%
\url{https://doi.org/10.1145/3447526.3472044}
\showURL{%
\tempurl}


\bibitem[\protect\citeauthoryear{Arefin~Shimon, Lutton, Xu, Morrison-Smith,
  Boucher, and Ruiz}{Arefin~Shimon et~al\mbox{.}}{2016}]%
        {arefin2016exploring}
\bibfield{author}{\bibinfo{person}{Shaikh~Shawon Arefin~Shimon},
  \bibinfo{person}{Courtney Lutton}, \bibinfo{person}{Zichun Xu},
  \bibinfo{person}{Sarah Morrison-Smith}, \bibinfo{person}{Christina Boucher},
  {and} \bibinfo{person}{Jaime Ruiz}.} \bibinfo{year}{2016}\natexlab{}.
\newblock \showarticletitle{Exploring non-touchscreen gestures for
  smartwatches}. In \bibinfo{booktitle}{\emph{Proceedings of the 2016 chi
  conference on human factors in computing systems}}.
  \bibinfo{pages}{3822--3833}.
\newblock


\bibitem[\protect\citeauthoryear{Ascari, Pereira, and Silva}{Ascari
  et~al\mbox{.}}{2020}]%
        {10.1145/3408300}
\bibfield{author}{\bibinfo{person}{R\'{u}bia E. O.~Schultz Ascari},
  \bibinfo{person}{Roberto Pereira}, {and} \bibinfo{person}{Luciano Silva}.}
  \bibinfo{year}{2020}\natexlab{}.
\newblock \showarticletitle{Computer Vision-Based Methodology to Improve
  Interaction for People with Motor and Speech Impairment}.
\newblock \bibinfo{journal}{\emph{ACM Trans. Access. Comput.}}
  \bibinfo{volume}{13}, \bibinfo{number}{4}, Article \bibinfo{articleno}{14}
  (\bibinfo{date}{Oct.} \bibinfo{year}{2020}), \bibinfo{numpages}{33}~pages.
\newblock
\showISSN{1936-7228}
\urldef\tempurl%
\url{https://doi.org/10.1145/3408300}
\showDOI{\tempurl}


\bibitem[\protect\citeauthoryear{Blain-Moraes, Schaff, Gruis, Huggins, and
  Wren}{Blain-Moraes et~al\mbox{.}}{2012}]%
        {blain2012barriers}
\bibfield{author}{\bibinfo{person}{Stefanie Blain-Moraes},
  \bibinfo{person}{Riley Schaff}, \bibinfo{person}{Kirsten~L Gruis},
  \bibinfo{person}{Jane~E Huggins}, {and} \bibinfo{person}{Patricia~A Wren}.}
  \bibinfo{year}{2012}\natexlab{}.
\newblock \showarticletitle{Barriers to and mediators of brain--computer
  interface user acceptance: focus group findings}.
\newblock \bibinfo{journal}{\emph{Ergonomics}} \bibinfo{volume}{55},
  \bibinfo{number}{5} (\bibinfo{year}{2012}), \bibinfo{pages}{516--525}.
\newblock


\bibitem[\protect\citeauthoryear{Corralejo, Nicol{\'a}s-Alonso, {\'A}lvarez,
  and Hornero}{Corralejo et~al\mbox{.}}{2014}]%
        {corralejo2014p300}
\bibfield{author}{\bibinfo{person}{Rebeca Corralejo}, \bibinfo{person}{Luis~F
  Nicol{\'a}s-Alonso}, \bibinfo{person}{Daniel {\'A}lvarez}, {and}
  \bibinfo{person}{Roberto Hornero}.} \bibinfo{year}{2014}\natexlab{}.
\newblock \showarticletitle{A P300-based brain--computer interface aimed at
  operating electronic devices at home for severely disabled people}.
\newblock \bibinfo{journal}{\emph{Medical \& biological engineering \&
  computing}} \bibinfo{volume}{52}, \bibinfo{number}{10}
  (\bibinfo{year}{2014}), \bibinfo{pages}{861--872}.
\newblock


\bibitem[\protect\citeauthoryear{Dingler, Rzayev, Shirazi, and Henze}{Dingler
  et~al\mbox{.}}{2018}]%
        {dingler2018designing}
\bibfield{author}{\bibinfo{person}{Tilman Dingler}, \bibinfo{person}{Rufat
  Rzayev}, \bibinfo{person}{Alireza~Sahami Shirazi}, {and}
  \bibinfo{person}{Niels Henze}.} \bibinfo{year}{2018}\natexlab{}.
\newblock \showarticletitle{Designing consistent gestures across device types:
  Eliciting RSVP controls for phone, watch, and glasses}. In
  \bibinfo{booktitle}{\emph{Proceedings of the 2018 CHI Conference on Human
  Factors in Computing Systems}}. \bibinfo{pages}{1--12}.
\newblock


\bibitem[\protect\citeauthoryear{Dong, Piumsomboon, Zhang, Clark, Bai, and
  Lindeman}{Dong et~al\mbox{.}}{2020a}]%
        {10.1145/3334480.3382883}
\bibfield{author}{\bibinfo{person}{Ze Dong}, \bibinfo{person}{Thammathip
  Piumsomboon}, \bibinfo{person}{Jingjing Zhang}, \bibinfo{person}{Adrian
  Clark}, \bibinfo{person}{Huidong Bai}, {and} \bibinfo{person}{Rob Lindeman}.}
  \bibinfo{year}{2020}\natexlab{a}.
\newblock \showarticletitle{A Comparison of Surface and Motion User-Defined
  Gestures for Mobile Augmented Reality}. In \bibinfo{booktitle}{\emph{Extended
  Abstracts of the 2020 CHI Conference on Human Factors in Computing Systems}}
  (Honolulu, HI, USA) \emph{(\bibinfo{series}{CHI EA '20})}.
  \bibinfo{publisher}{Association for Computing Machinery},
  \bibinfo{address}{New York, NY, USA}, \bibinfo{pages}{1–8}.
\newblock
\showISBNx{9781450368193}
\urldef\tempurl%
\url{https://doi.org/10.1145/3334480.3382883}
\showDOI{\tempurl}


\bibitem[\protect\citeauthoryear{Dong, Zhang, Lindeman, and Piumsomboon}{Dong
  et~al\mbox{.}}{2020b}]%
        {10.1145/3385959.3422694}
\bibfield{author}{\bibinfo{person}{Ze Dong}, \bibinfo{person}{Jingjing Zhang},
  \bibinfo{person}{Robert Lindeman}, {and} \bibinfo{person}{Thammathip
  Piumsomboon}.} \bibinfo{year}{2020}\natexlab{b}.
\newblock \showarticletitle{Surface vs Motion Gestures for Mobile Augmented
  Reality}. In \bibinfo{booktitle}{\emph{Symposium on Spatial User
  Interaction}} (Virtual Event, Canada) \emph{(\bibinfo{series}{SUI '20})}.
  \bibinfo{publisher}{Association for Computing Machinery},
  \bibinfo{address}{New York, NY, USA}, Article \bibinfo{articleno}{30},
  \bibinfo{numpages}{2}~pages.
\newblock
\showISBNx{9781450379434}
\urldef\tempurl%
\url{https://doi.org/10.1145/3385959.3422694}
\showDOI{\tempurl}


\bibitem[\protect\citeauthoryear{Fajardo-Flores, Gayt\'{a}n-Lugo,
  Santana-Mancilla, and Rodr\'{\i}guez-Ortiz}{Fajardo-Flores
  et~al\mbox{.}}{2017}]%
        {10.1145/3151470.3151476}
\bibfield{author}{\bibinfo{person}{Silvia~B. Fajardo-Flores},
  \bibinfo{person}{Laura~S. Gayt\'{a}n-Lugo}, \bibinfo{person}{Pedro~C.
  Santana-Mancilla}, {and} \bibinfo{person}{Miguel~A. Rodr\'{\i}guez-Ortiz}.}
  \bibinfo{year}{2017}\natexlab{}.
\newblock \showarticletitle{Mobile Accessibility for People with Combined
  Visual and Motor Impairment: A Case Study}. In
  \bibinfo{booktitle}{\emph{Proceedings of the 8th Latin American Conference on
  Human-Computer Interaction}} (Antigua Guatemala, Guatemala)
  \emph{(\bibinfo{series}{CLIHC '17})}. \bibinfo{publisher}{Association for
  Computing Machinery}, \bibinfo{address}{New York, NY, USA}, Article
  \bibinfo{articleno}{7}, \bibinfo{numpages}{4}~pages.
\newblock
\showISBNx{9781450354295}
\urldef\tempurl%
\url{https://doi.org/10.1145/3151470.3151476}
\showDOI{\tempurl}


\bibitem[\protect\citeauthoryear{Fan, Li, and Li}{Fan et~al\mbox{.}}{2020}]%
        {10.1145/3373625.3416987}
\bibfield{author}{\bibinfo{person}{Mingming Fan}, \bibinfo{person}{Zhen Li},
  {and} \bibinfo{person}{Franklin~Mingzhe Li}.}
  \bibinfo{year}{2020}\natexlab{}.
\newblock \showarticletitle{Eyelid Gestures on Mobile Devices for People with
  Motor Impairments}. In \bibinfo{booktitle}{\emph{The 22nd International ACM
  SIGACCESS Conference on Computers and Accessibility}} (Virtual Event, Greece)
  \emph{(\bibinfo{series}{ASSETS '20})}. \bibinfo{publisher}{Association for
  Computing Machinery}, \bibinfo{address}{New York, NY, USA}, Article
  \bibinfo{articleno}{15}, \bibinfo{numpages}{8}~pages.
\newblock
\showISBNx{9781450371032}
\urldef\tempurl%
\url{https://doi.org/10.1145/3373625.3416987}
\showDOI{\tempurl}


\bibitem[\protect\citeauthoryear{Fan, Li, and Li}{Fan et~al\mbox{.}}{2021}]%
        {fan2021eyelidCAM}
\bibfield{author}{\bibinfo{person}{Mingming Fan}, \bibinfo{person}{Zhen Li},
  {and} \bibinfo{person}{Franklin~Mingzhe Li}.}
  \bibinfo{year}{2021}\natexlab{}.
\newblock \showarticletitle{Eyelid gestures for people with motor impairments}.
\newblock \bibinfo{journal}{\emph{Commun. ACM}} \bibinfo{volume}{65},
  \bibinfo{number}{1} (\bibinfo{year}{2021}), \bibinfo{pages}{108--115}.
\newblock


\bibitem[\protect\citeauthoryear{Heikkil{\"{a}} and
  R{\"{a}}ih{\"{a}}}{Heikkil{\"{a}} and R{\"{a}}ih{\"{a}}}{2012}]%
        {Heikkila2012}
\bibfield{author}{\bibinfo{person}{Henna Heikkil{\"{a}}} {and}
  \bibinfo{person}{Kari-Jouko R{\"{a}}ih{\"{a}}}.}
  \bibinfo{year}{2012}\natexlab{}.
\newblock \showarticletitle{{Simple gaze gestures and the closure of the eyes
  as an interaction technique}}. In \bibinfo{booktitle}{\emph{Proceedings of
  the Symposium on Eye Tracking Research and Applications - ETRA '12}}.
  \bibinfo{publisher}{ACM Press}, \bibinfo{address}{New York, New York, USA},
  \bibinfo{pages}{147}.
\newblock
\showISBNx{9781450312219}
\urldef\tempurl%
\url{https://doi.org/10.1145/2168556.2168579}
\showDOI{\tempurl}


\bibitem[\protect\citeauthoryear{Ishimaru, Kunze, Kise, Weppner, Dengel,
  Lukowicz, and Bulling}{Ishimaru et~al\mbox{.}}{2014}]%
        {Ishimaru2014}
\bibfield{author}{\bibinfo{person}{Shoya Ishimaru}, \bibinfo{person}{Kai
  Kunze}, \bibinfo{person}{Koichi Kise}, \bibinfo{person}{Jens Weppner},
  \bibinfo{person}{Andreas Dengel}, \bibinfo{person}{Paul Lukowicz}, {and}
  \bibinfo{person}{Andreas Bulling}.} \bibinfo{year}{2014}\natexlab{}.
\newblock \showarticletitle{In the Blink of an Eye: Combining Head Motion and
  Eye Blink Frequency for Activity Recognition with Google Glass}. In
  \bibinfo{booktitle}{\emph{Proceedings of the 5th Augmented Human
  International Conference}} (Kobe, Japan) \emph{(\bibinfo{series}{AH '14})}.
  \bibinfo{publisher}{ACM}, \bibinfo{address}{New York, NY, USA}, Article
  \bibinfo{articleno}{15}, \bibinfo{numpages}{4}~pages.
\newblock
\showISBNx{978-1-4503-2761-9}
\urldef\tempurl%
\url{https://doi.org/10.1145/2582051.2582066}
\showDOI{\tempurl}


\bibitem[\protect\citeauthoryear{Kaufman, Bandopadhay, and Shaviv}{Kaufman
  et~al\mbox{.}}{1993}]%
        {kaufman1993}
\bibfield{author}{\bibinfo{person}{Arie~E Kaufman}, \bibinfo{person}{Amit
  Bandopadhay}, {and} \bibinfo{person}{Bernard~D Shaviv}.}
  \bibinfo{year}{1993}\natexlab{}.
\newblock \showarticletitle{An eye tracking computer user interface}. In
  \bibinfo{booktitle}{\emph{Proceedings of 1993 IEEE Research Properties in
  Virtual Reality Symposium}}. IEEE, \bibinfo{pages}{120--121}.
\newblock


\bibitem[\protect\citeauthoryear{Kurdyukova, Redlin, and Andr\'{e}}{Kurdyukova
  et~al\mbox{.}}{2012}]%
        {10.1145/2166966.2166984}
\bibfield{author}{\bibinfo{person}{Ekaterina Kurdyukova},
  \bibinfo{person}{Matthias Redlin}, {and} \bibinfo{person}{Elisabeth
  Andr\'{e}}.} \bibinfo{year}{2012}\natexlab{}.
\newblock \showarticletitle{Studying User-Defined IPad Gestures for Interaction
  in Multi-Display Environment}. In \bibinfo{booktitle}{\emph{Proceedings of
  the 2012 ACM International Conference on Intelligent User Interfaces}}
  (Lisbon, Portugal) \emph{(\bibinfo{series}{IUI '12})}.
  \bibinfo{publisher}{Association for Computing Machinery},
  \bibinfo{address}{New York, NY, USA}, \bibinfo{pages}{93–96}.
\newblock
\showISBNx{9781450310482}
\urldef\tempurl%
\url{https://doi.org/10.1145/2166966.2166984}
\showDOI{\tempurl}


\bibitem[\protect\citeauthoryear{Kwon and Kim}{Kwon and Kim}{1999}]%
        {Kwon1999}
\bibfield{author}{\bibinfo{person}{S.H. Kwon} {and} \bibinfo{person}{H.C.
  Kim}.} \bibinfo{year}{1999}\natexlab{}.
\newblock \showarticletitle{{EOG-based glasses-type wireless mouse for the
  disabled}}.
\newblock \bibinfo{journal}{\emph{Proceedings of the First Joint BMES/EMBS
  Conference. 1999 IEEE Engineering in Medicine and Biology 21st Annual
  Conference and the 1999 Annual Fall Meeting of the Biomedical Engineering
  Society (Cat. No.99CH37015)}}  \bibinfo{volume}{1} (\bibinfo{year}{1999}),
  \bibinfo{pages}{592}.
\newblock
\showISBNx{0-7803-5674-8}
\showISSN{1094-687X}
\urldef\tempurl%
\url{https://doi.org/10.1109/IEMBS.1999.802670}
\showDOI{\tempurl}


\bibitem[\protect\citeauthoryear{Kyt{\"o}, Ens, Piumsomboon, Lee, and
  Billinghurst}{Kyt{\"o} et~al\mbox{.}}{2018}]%
        {kyto2018pinpointing}
\bibfield{author}{\bibinfo{person}{Mikko Kyt{\"o}}, \bibinfo{person}{Barrett
  Ens}, \bibinfo{person}{Thammathip Piumsomboon}, \bibinfo{person}{Gun~A Lee},
  {and} \bibinfo{person}{Mark Billinghurst}.} \bibinfo{year}{2018}\natexlab{}.
\newblock \showarticletitle{Pinpointing: Precise head-and eye-based target
  selection for augmented reality}. In \bibinfo{booktitle}{\emph{Proceedings of
  the 2018 CHI Conference on Human Factors in Computing Systems}}.
  \bibinfo{pages}{1--14}.
\newblock


\bibitem[\protect\citeauthoryear{Lazarou, Nikolopoulos, Petrantonakis,
  Kompatsiaris, and Tsolaki}{Lazarou et~al\mbox{.}}{2018}]%
        {10.3389/fnhum.2018.00014}
\bibfield{author}{\bibinfo{person}{Ioulietta Lazarou}, \bibinfo{person}{Spiros
  Nikolopoulos}, \bibinfo{person}{Panagiotis~C. Petrantonakis},
  \bibinfo{person}{Ioannis Kompatsiaris}, {and} \bibinfo{person}{Magda
  Tsolaki}.} \bibinfo{year}{2018}\natexlab{}.
\newblock \showarticletitle{EEG-Based Brain–Computer Interfaces for
  Communication and Rehabilitation of People with Motor Impairment: A Novel
  Approach of the 21st Century}.
\newblock \bibinfo{journal}{\emph{Frontiers in Human Neuroscience}}
  \bibinfo{volume}{12} (\bibinfo{year}{2018}), \bibinfo{pages}{14}.
\newblock
\showISSN{1662-5161}
\urldef\tempurl%
\url{https://doi.org/10.3389/fnhum.2018.00014}
\showDOI{\tempurl}


\bibitem[\protect\citeauthoryear{Lee, Wong, Park, Choi, Park, and
  Billinghurst}{Lee et~al\mbox{.}}{2015}]%
        {10.1145/2702613.2732747}
\bibfield{author}{\bibinfo{person}{Gun~A. Lee}, \bibinfo{person}{Jonathan
  Wong}, \bibinfo{person}{Hye~Sun Park}, \bibinfo{person}{Jin~Sung Choi},
  \bibinfo{person}{Chang~Joon Park}, {and} \bibinfo{person}{Mark
  Billinghurst}.} \bibinfo{year}{2015}\natexlab{}.
\newblock \showarticletitle{User Defined Gestures for Augmented Virtual
  Mirrors: A Guessability Study}. In \bibinfo{booktitle}{\emph{Proceedings of
  the 33rd Annual ACM Conference Extended Abstracts on Human Factors in
  Computing Systems}} (Seoul, Republic of Korea) \emph{(\bibinfo{series}{CHI EA
  '15})}. \bibinfo{publisher}{Association for Computing Machinery},
  \bibinfo{address}{New York, NY, USA}, \bibinfo{pages}{959–964}.
\newblock
\showISBNx{9781450331463}
\urldef\tempurl%
\url{https://doi.org/10.1145/2702613.2732747}
\showDOI{\tempurl}


\bibitem[\protect\citeauthoryear{Li, Fan, Han, and Truong}{Li
  et~al\mbox{.}}{2020}]%
        {10.1145/3422852.3423479}
\bibfield{author}{\bibinfo{person}{Zhen Li}, \bibinfo{person}{Mingming Fan},
  \bibinfo{person}{Ying Han}, {and} \bibinfo{person}{Khai~N. Truong}.}
  \bibinfo{year}{2020}\natexlab{}.
\newblock \showarticletitle{IWink: Exploring Eyelid Gestures on Mobile
  Devices}. In \bibinfo{booktitle}{\emph{Proceedings of the 1st International
  Workshop on Human-Centric Multimedia Analysis}} (Seattle, WA, USA)
  \emph{(\bibinfo{series}{HuMA'20})}. \bibinfo{publisher}{Association for
  Computing Machinery}, \bibinfo{address}{New York, NY, USA},
  \bibinfo{pages}{83–89}.
\newblock
\showISBNx{9781450381512}
\urldef\tempurl%
\url{https://doi.org/10.1145/3422852.3423479}
\showDOI{\tempurl}


\bibitem[\protect\citeauthoryear{Malu and Findlater}{Malu and
  Findlater}{2015}]%
        {10.1145/2702123.2702188}
\bibfield{author}{\bibinfo{person}{Meethu Malu} {and} \bibinfo{person}{Leah
  Findlater}.} \bibinfo{year}{2015}\natexlab{}.
\newblock \bibinfo{booktitle}{\emph{Personalized, Wearable Control of a
  Head-Mounted Display for Users with Upper Body Motor Impairments}}.
\newblock \bibinfo{publisher}{Association for Computing Machinery},
  \bibinfo{address}{New York, NY, USA}, \bibinfo{pages}{221–230}.
\newblock
\showISBNx{9781450331456}
\urldef\tempurl%
\url{https://doi-org.ezproxy.rit.edu/10.1145/2702123.2702188}
\showURL{%
\tempurl}


\bibitem[\protect\citeauthoryear{Mott, E., Bennett, Cutrell, and Morris}{Mott
  et~al\mbox{.}}{2018}]%
        {10.1145/3173574.3174094}
\bibfield{author}{\bibinfo{person}{Martez~E. Mott}, \bibinfo{person}{Jane E.},
  \bibinfo{person}{Cynthia~L. Bennett}, \bibinfo{person}{Edward Cutrell}, {and}
  \bibinfo{person}{Meredith~Ringel Morris}.} \bibinfo{year}{2018}\natexlab{}.
\newblock \bibinfo{booktitle}{\emph{Understanding the Accessibility of
  Smartphone Photography for People with Motor Impairments}}.
\newblock \bibinfo{publisher}{Association for Computing Machinery},
  \bibinfo{address}{New York, NY, USA}, \bibinfo{pages}{1–12}.
\newblock
\showISBNx{9781450356206}
\urldef\tempurl%
\url{https://doi.org/10.1145/3173574.3174094}
\showURL{%
\tempurl}


\bibitem[\protect\citeauthoryear{Naftali and Findlater}{Naftali and
  Findlater}{2014}]%
        {10.1145/2661334.2661372}
\bibfield{author}{\bibinfo{person}{Maia Naftali} {and} \bibinfo{person}{Leah
  Findlater}.} \bibinfo{year}{2014}\natexlab{}.
\newblock \showarticletitle{Accessibility in Context: Understanding the Truly
  Mobile Experience of Smartphone Users with Motor Impairments}. In
  \bibinfo{booktitle}{\emph{Proceedings of the 16th International ACM SIGACCESS
  Conference on Computers and Accessibility}} (Rochester, New York, USA)
  \emph{(\bibinfo{series}{ASSETS '14})}. \bibinfo{publisher}{Association for
  Computing Machinery}, \bibinfo{address}{New York, NY, USA},
  \bibinfo{pages}{209–216}.
\newblock
\showISBNx{9781450327206}
\urldef\tempurl%
\url{https://doi.org/10.1145/2661334.2661372}
\showDOI{\tempurl}


\bibitem[\protect\citeauthoryear{Nukarinen, Kangas, {\v{S}}pakov, Isokoski,
  Akkil, Rantala, and Raisamo}{Nukarinen et~al\mbox{.}}{2016}]%
        {nukarinen2016evaluation}
\bibfield{author}{\bibinfo{person}{Tomi Nukarinen}, \bibinfo{person}{Jari
  Kangas}, \bibinfo{person}{Oleg {\v{S}}pakov}, \bibinfo{person}{Poika
  Isokoski}, \bibinfo{person}{Deepak Akkil}, \bibinfo{person}{Jussi Rantala},
  {and} \bibinfo{person}{Roope Raisamo}.} \bibinfo{year}{2016}\natexlab{}.
\newblock \showarticletitle{Evaluation of HeadTurn: An interaction technique
  using the gaze and head turns}. In \bibinfo{booktitle}{\emph{Proceedings of
  the 9th Nordic Conference on Human-Computer Interaction}}.
  \bibinfo{pages}{1--8}.
\newblock


\bibitem[\protect\citeauthoryear{Pires, Nunes, and Castelo-Branco}{Pires
  et~al\mbox{.}}{2012}]%
        {pires2012evaluation}
\bibfield{author}{\bibinfo{person}{Gabriel Pires}, \bibinfo{person}{Urbano
  Nunes}, {and} \bibinfo{person}{Miguel Castelo-Branco}.}
  \bibinfo{year}{2012}\natexlab{}.
\newblock \showarticletitle{Evaluation of brain-computer interfaces in
  accessing computer and other devices by people with severe motor
  impairments}.
\newblock \bibinfo{journal}{\emph{Procedia Computer Science}}
  \bibinfo{volume}{14} (\bibinfo{year}{2012}), \bibinfo{pages}{283--292}.
\newblock


\bibitem[\protect\citeauthoryear{Piumsomboon, Clark, Billinghurst, and
  Cockburn}{Piumsomboon et~al\mbox{.}}{2013}]%
        {10.1145/2468356.2468527}
\bibfield{author}{\bibinfo{person}{Thammathip Piumsomboon},
  \bibinfo{person}{Adrian Clark}, \bibinfo{person}{Mark Billinghurst}, {and}
  \bibinfo{person}{Andy Cockburn}.} \bibinfo{year}{2013}\natexlab{}.
\newblock \showarticletitle{User-Defined Gestures for Augmented Reality}. In
  \bibinfo{booktitle}{\emph{CHI '13 Extended Abstracts on Human Factors in
  Computing Systems}} (Paris, France) \emph{(\bibinfo{series}{CHI EA '13})}.
  \bibinfo{publisher}{Association for Computing Machinery},
  \bibinfo{address}{New York, NY, USA}, \bibinfo{pages}{955–960}.
\newblock
\showISBNx{9781450319522}
\urldef\tempurl%
\url{https://doi.org/10.1145/2468356.2468527}
\showDOI{\tempurl}


\bibitem[\protect\citeauthoryear{Ruiz, Li, and Lank}{Ruiz
  et~al\mbox{.}}{2011}]%
        {10.1145/1978942.1978971}
\bibfield{author}{\bibinfo{person}{Jaime Ruiz}, \bibinfo{person}{Yang Li},
  {and} \bibinfo{person}{Edward Lank}.} \bibinfo{year}{2011}\natexlab{}.
\newblock \bibinfo{booktitle}{\emph{User-Defined Motion Gestures for Mobile
  Interaction}}.
\newblock \bibinfo{publisher}{Association for Computing Machinery},
  \bibinfo{address}{New York, NY, USA}, \bibinfo{pages}{197–206}.
\newblock
\showISBNx{9781450302289}
\urldef\tempurl%
\url{https://doi.org/10.1145/1978942.1978971}
\showURL{%
\tempurl}


\bibitem[\protect\citeauthoryear{Shaw, Crisman, Loomis, and Laszewski}{Shaw
  et~al\mbox{.}}{1990}]%
        {Shaw1990}
\bibfield{author}{\bibinfo{person}{Robin Shaw}, \bibinfo{person}{Everett
  Crisman}, \bibinfo{person}{Anne Loomis}, {and} \bibinfo{person}{Zofia
  Laszewski}.} \bibinfo{year}{1990}\natexlab{}.
\newblock \showarticletitle{{The eye wink control interface: using the computer
  to provide the severely disabled with increased flexibility and comfort}}. In
  \bibinfo{booktitle}{\emph{Proc. of the Third Annual IEEE Symposium on
  Computer-Based Medical Systems}}. \bibinfo{publisher}{IEEE},
  \bibinfo{pages}{105--111}.
\newblock
\showISBNx{0-8186-9040-2}
\urldef\tempurl%
\url{https://doi.org/10.1109/CBMSYS.1990.109386}
\showDOI{\tempurl}


\bibitem[\protect\citeauthoryear{Sidenmark and Gellersen}{Sidenmark and
  Gellersen}{2019}]%
        {sidenmark2019eye}
\bibfield{author}{\bibinfo{person}{Ludwig Sidenmark} {and}
  \bibinfo{person}{Hans Gellersen}.} \bibinfo{year}{2019}\natexlab{}.
\newblock \showarticletitle{Eye\&head: Synergetic eye and head movement for
  gaze pointing and selection}. In \bibinfo{booktitle}{\emph{Proceedings of the
  32nd Annual ACM Symposium on User Interface Software and Technology}}.
  \bibinfo{pages}{1161--1174}.
\newblock


\bibitem[\protect\citeauthoryear{Taherian, Selitskiy, Pau, and
  Claire~Davies}{Taherian et~al\mbox{.}}{2017}]%
        {taherian2017we}
\bibfield{author}{\bibinfo{person}{Sarvnaz Taherian}, \bibinfo{person}{Dmitry
  Selitskiy}, \bibinfo{person}{James Pau}, {and} \bibinfo{person}{T
  Claire~Davies}.} \bibinfo{year}{2017}\natexlab{}.
\newblock \showarticletitle{Are we there yet? Evaluating commercial grade
  brain--computer interface for control of computer applications by individuals
  with cerebral palsy}.
\newblock \bibinfo{journal}{\emph{Disability and Rehabilitation: Assistive
  Technology}} \bibinfo{volume}{12}, \bibinfo{number}{2}
  (\bibinfo{year}{2017}), \bibinfo{pages}{165--174}.
\newblock


\bibitem[\protect\citeauthoryear{Troiano, Pedersen, and Hornb\ae{}k}{Troiano
  et~al\mbox{.}}{2014}]%
        {10.1145/2598153.2598184}
\bibfield{author}{\bibinfo{person}{Giovanni~Maria Troiano},
  \bibinfo{person}{Esben~Warming Pedersen}, {and} \bibinfo{person}{Kasper
  Hornb\ae{}k}.} \bibinfo{year}{2014}\natexlab{}.
\newblock \showarticletitle{User-Defined Gestures for Elastic, Deformable
  Displays}. In \bibinfo{booktitle}{\emph{Proceedings of the 2014 International
  Working Conference on Advanced Visual Interfaces}} (Como, Italy)
  \emph{(\bibinfo{series}{AVI '14})}. \bibinfo{publisher}{Association for
  Computing Machinery}, \bibinfo{address}{New York, NY, USA},
  \bibinfo{pages}{1–8}.
\newblock
\showISBNx{9781450327756}
\urldef\tempurl%
\url{https://doi.org/10.1145/2598153.2598184}
\showDOI{\tempurl}


\bibitem[\protect\citeauthoryear{Ungurean, Vatavu, Leiva, and
  Mart\'{\i}n-Albo}{Ungurean et~al\mbox{.}}{2018}]%
        {10.1145/3236112.3236116}
\bibfield{author}{\bibinfo{person}{Ovidiu-Ciprian Ungurean},
  \bibinfo{person}{Radu-Daniel Vatavu}, \bibinfo{person}{Luis~A. Leiva}, {and}
  \bibinfo{person}{Daniel Mart\'{\i}n-Albo}.} \bibinfo{year}{2018}\natexlab{}.
\newblock \showarticletitle{Predicting Stroke Gesture Input Performance for
  Users with Motor Impairments}. In \bibinfo{booktitle}{\emph{Proceedings of
  the 20th International Conference on Human-Computer Interaction with Mobile
  Devices and Services Adjunct}} (Barcelona, Spain)
  \emph{(\bibinfo{series}{MobileHCI '18})}. \bibinfo{publisher}{Association for
  Computing Machinery}, \bibinfo{address}{New York, NY, USA},
  \bibinfo{pages}{23–30}.
\newblock
\showISBNx{9781450359412}
\urldef\tempurl%
\url{https://doi.org/10.1145/3236112.3236116}
\showDOI{\tempurl}


\bibitem[\protect\citeauthoryear{Vatavu and Wobbrock}{Vatavu and
  Wobbrock}{2015}]%
        {vatavu2015formalizing}
\bibfield{author}{\bibinfo{person}{Radu-Daniel Vatavu} {and}
  \bibinfo{person}{Jacob~O Wobbrock}.} \bibinfo{year}{2015}\natexlab{}.
\newblock \showarticletitle{Formalizing agreement analysis for elicitation
  studies: new measures, significance test, and toolkit}. In
  \bibinfo{booktitle}{\emph{Proceedings of the 33rd Annual ACM Conference on
  Human Factors in Computing Systems}}. \bibinfo{pages}{1325--1334}.
\newblock


\bibitem[\protect\citeauthoryear{Weidner and Broll}{Weidner and Broll}{2019}]%
        {10.1145/3365610.3365625}
\bibfield{author}{\bibinfo{person}{Florian Weidner} {and}
  \bibinfo{person}{Wolfgang Broll}.} \bibinfo{year}{2019}\natexlab{}.
\newblock \showarticletitle{Interact with Your Car: A User-Elicited Gesture Set
  to Inform Future in-Car User Interfaces}. In
  \bibinfo{booktitle}{\emph{Proceedings of the 18th International Conference on
  Mobile and Ubiquitous Multimedia}} (Pisa, Italy) \emph{(\bibinfo{series}{MUM
  '19})}. \bibinfo{publisher}{Association for Computing Machinery},
  \bibinfo{address}{New York, NY, USA}, Article \bibinfo{articleno}{11},
  \bibinfo{numpages}{12}~pages.
\newblock
\showISBNx{9781450376242}
\urldef\tempurl%
\url{https://doi.org/10.1145/3365610.3365625}
\showDOI{\tempurl}


\bibitem[\protect\citeauthoryear{Wobbrock, Aung, Rothrock, and Myers}{Wobbrock
  et~al\mbox{.}}{2005}]%
        {wobbrock2005maximizing}
\bibfield{author}{\bibinfo{person}{Jacob~O Wobbrock},
  \bibinfo{person}{Htet~Htet Aung}, \bibinfo{person}{Brandon Rothrock}, {and}
  \bibinfo{person}{Brad~A Myers}.} \bibinfo{year}{2005}\natexlab{}.
\newblock \showarticletitle{Maximizing the guessability of symbolic input}. In
  \bibinfo{booktitle}{\emph{CHI'05 extended abstracts on Human Factors in
  Computing Systems}}. \bibinfo{pages}{1869--1872}.
\newblock


\bibitem[\protect\citeauthoryear{Wobbrock, Morris, and Wilson}{Wobbrock
  et~al\mbox{.}}{2009}]%
        {10.1145/1518701.1518866}
\bibfield{author}{\bibinfo{person}{Jacob~O. Wobbrock},
  \bibinfo{person}{Meredith~Ringel Morris}, {and} \bibinfo{person}{Andrew~D.
  Wilson}.} \bibinfo{year}{2009}\natexlab{}.
\newblock \bibinfo{booktitle}{\emph{User-Defined Gestures for Surface
  Computing}}.
\newblock \bibinfo{publisher}{Association for Computing Machinery},
  \bibinfo{address}{New York, NY, USA}, \bibinfo{pages}{1083–1092}.
\newblock
\showISBNx{9781605582467}
\urldef\tempurl%
\url{https://doi.org/10.1145/1518701.1518866}
\showURL{%
\tempurl}


\bibitem[\protect\citeauthoryear{Yan, Yu, Yi, and Shi}{Yan
  et~al\mbox{.}}{2018}]%
        {yan2018headgesture}
\bibfield{author}{\bibinfo{person}{Yukang Yan}, \bibinfo{person}{Chun Yu},
  \bibinfo{person}{Xin Yi}, {and} \bibinfo{person}{Yuanchun Shi}.}
  \bibinfo{year}{2018}\natexlab{}.
\newblock \showarticletitle{Headgesture: Hands-free input approach leveraging
  head movements for hmd devices}.
\newblock \bibinfo{journal}{\emph{Proceedings of the ACM on Interactive,
  Mobile, Wearable and Ubiquitous Technologies}} \bibinfo{volume}{2},
  \bibinfo{number}{4} (\bibinfo{year}{2018}), \bibinfo{pages}{1--23}.
\newblock


\bibitem[\protect\citeauthoryear{Zhang, Kulkarni, and Morris}{Zhang
  et~al\mbox{.}}{2017}]%
        {Zhang:2017:SGG:3025453.3025790}
\bibfield{author}{\bibinfo{person}{Xiaoyi Zhang}, \bibinfo{person}{Harish
  Kulkarni}, {and} \bibinfo{person}{Meredith~Ringel Morris}.}
  \bibinfo{year}{2017}\natexlab{}.
\newblock \showarticletitle{Smartphone-Based Gaze Gesture Communication for
  People with Motor Disabilities}. In \bibinfo{booktitle}{\emph{Proceedings of
  the 2017 CHI Conference on Human Factors in Computing Systems}} (Denver,
  Colorado, USA) \emph{(\bibinfo{series}{CHI '17})}. \bibinfo{publisher}{ACM},
  \bibinfo{address}{New York, NY, USA}, \bibinfo{pages}{2878--2889}.
\newblock
\showISBNx{978-1-4503-4655-9}
\urldef\tempurl%
\url{https://doi.org/10.1145/3025453.3025790}
\showDOI{\tempurl}


\end{thebibliography}


\end{document}